\documentclass[twocolumn, times]{aastex631}

\usepackage{xspace}
\newcommand{\Ntwo}{N$_2$\xspace}
\newcommand{\ammonia}{NH$_3$\xspace}
\newcommand{\dammonia}{NH$_2$D\xspace}
\newcommand{\methanol}{CH$_3$OH\xspace}
\newcommand{\water}{H$_2$O\xspace}
\newcommand{\nthp}{N$_2$H$^+$\xspace}
\newcommand{\DDratio}{[NHD$_2$/NH$_2$D]/[NH$_2$D/NH$_3$]\xspace}

\shorttitle{\dammonia/\ammonia in IRAS 4A}
\shortauthors{Yamato et al.}
\graphicspath{{./}{figures/}}

\begin{document}

\title{The first interferometric measurements of \dammonia/\ammonia ratio in hot corinos}

\author[0000-0003-4099-6941]{Yoshihide Yamato}
\affil{Department of Astronomy, Graduate School of Science, The University of Tokyo, 7-3-1 Hongo, Bunkyo-ku, Tokyo 113-0033, Japan}

\author[0000-0002-2026-8157]{Kenji Furuya}
\affil{National Astronomical Observatory of Japan, Osawa 2-21-1, Mitaka, Tokyo 181-8588, Japan}

\author[0000-0003-3283-6884]{Yuri Aikawa}
\affil{Department of Astronomy, Graduate School of Science, The University of Tokyo, 7-3-1 Hongo, Bunkyo-ku, Tokyo 113-0033, Japan}

\author[0000-0002-1100-5734]{Magnus V. Persson}
\affil{Department of Space, Earth and Environment, Chalmers University of Technology, Onsala Space Observatory, 43992 Onsala, Sweden}

\author[0000-0002-6195-0152]{John J. Tobin}
\affil{National Radio Astronomy Observatory, 520 Edgemont Rd., Charlottesville, VA 22093 USA}

\author[0000-0001-9133-8047]{Jes K. J{\o}rgensen}
\affil{Niels Bohr Institute, University of Copenhagen, {\O}ster Voldgade 5--7, DK-1350 Copenhagen K., Denmark}

\author[0000-0003-0065-7267]{Mihkel Kama}
\affil{Department of Physics and Astronomy, University College London, Gower Street, WC1E 6BT London, United Kingdom}
\affil{Tartu Observatory, University of Tartu, 61602 Tõravere, Tartumaa, Estonia}



\begin{abstract}
The nitrogen chemical evolution during star and planet formation is still not fully understood. Ammonia (NH$_3$) is a key specie in the understanding of the molecular evolution in star-forming clouds and nitrogen isotope fractionation. In this paper, we present high spatial resolution observations of multiple emission lines of \ammonia toward the protobinary system NGC1333 IRAS4A with Karl G. Jansky Very Large Array (VLA). We spatially resolved the binary (hereafter 4A1 and 4A2) and detected compact emission of NH$_3$ transitions with high excitation energies ($\gtrsim$100\,K) from the vicinity of the protostars, indicating the NH$_3$ ice has sublimated at the inner hot region. The NH$_3$ column density is estimated to be $\sim$10$^{17}$--10$^{18}$\,cm$^{-2}$. We also detected two NH$_2$D transitions, allowing us to constrain the deuterium fractionation of ammonia. The NH$_2$D/NH$_3$ ratios are as high as $\sim$\textrm{0.3}--1 in both 4A1 and 4A2. From the comparisons with the astrochemical models in the literature, \textrm{the high \dammonia/\ammonia ratios} suggest that the formation of \ammonia ices mainly started in the prestellar phase after the formation of bulk water ice finished, and that the primary nitrogen reservoir in the star-forming cloud could be atomic nitrogen (or N atoms) rather than nitrogen-bearing species such as \Ntwo and \ammonia. The implications on the physical properties of IRAS4A cores are discussed as well. 


\end{abstract}

\keywords{}


\section{Introduction} \label{sec:intro}

Nitrogen is the fifth most abundant element in the interstellar medium (ISM) with an abundance of $\sim6\times10^{-5}$ with respect to hydrogen \citep[][]{Przybilla08}. Determining the main nitrogen reservoirs in molecular clouds is a fundamental problem in astrochemistry. The nitrogen budget in \textrm{molecular} clouds also affects the formation of nitrogen-bearing complex organics molecules and chemistry in planet-forming disks \citep[e.g.,][]{Schwarz14}.  

In \textrm{molecular} clouds, the main nitrogen reservoirs in the gas phase are expected to be molecular nitrogen (\Ntwo), which are formed in the gas-phase \citep[e.g.,][]{VanDishoeck93, Furuya18_L1544} \citep[see also][]{Oberg21}. While in the central region of the \textrm{clouds}, \Ntwo is self-shielded against the interstellar UV photons and thus abundant, a substantial amount of atomic nitrogen (or N atoms) could also be present in the outer regions where the density is relatively low. N and \Ntwo freeze-out onto the surface of dust grains in the \textrm{molecular} clouds, and subsequently form other nitrogen-bearing molecules such as ammonia (\ammonia) via a sequence of hydrogenation \textrm{\citep{Jonusas2020, Hidaka2011, Fedoseev2015}. Nitrogen-bearing molecules are also formed via gas-phase reactions \citep[e.g.,][]{LeGal2014}, and can successively deplete onto the dust grain mantles \citep[e.g.,][]{Caselli2022, Pineda2022}}. However, the partition of elemental nitrogen into these species is still not well constrained \citep{Oberg21}.   

Observationally constraining the main nitrogen reservoirs \textrm{is} generally difficult. 
Neither \Ntwo nor atomic N in the gas-phase can be observed due to the lack of low energy transitions that can be excited at the low temperature in \textrm{molecular} clouds. Instead, \citet{Maret06} used the \nthp emission line in \textrm{molecular} cloud cores to constrain the \Ntwo abundance in the gas phase. \nthp is primary formed by \Ntwo + H$_3^+$, and thus its abundance reflects the abundance of gaseous \Ntwo. Based on the relatively weak \nthp emission, they suggested that \Ntwo is not the main nitrogen reservoir, and instead atomic N in the gas phase would be. This is also consistent with the low \Ntwo abundance in comets. \citet{Rubin15} made the in situ measurement of \Ntwo in the comet 67P/Churyumov-Gerasimenko, revealing that the \Ntwo/CO ratio is depleted by a factor of $\sim$25 compared with the case where all protosolar C and N are in CO and \Ntwo. Icy nitrogen-bearing species \textrm{in molecular clouds} have been observed in infrared. \textrm{While the \ammonia ice absorption bands at 3\,\textmu m band have long been debated, its 9.1\,\textmu m band was clearly detected by \citet[][see also \citealt{Gibb2000}]{Lacy1998}}. Observations with the \textit{Spitzer} Space Telescope revealed that $\sim$10\,\% of overall nitrogen is locked up in ices, mainly as \ammonia \citep[e.g.,][]{Bottinelli10, Oberg11} \citep[see also][]{Boogert15}. Still, this relatively low \ammonia abundances imply that \Ntwo and N can be the main nitrogen reservoir in \textrm{molecular} clouds. Alternatively, there are also possibilities that a substantial amount of \ammonia is converted into ammonium salts, and they can constitute a significant portion of the nitrogen reservoir as observed in the comet 67P \citep{Altwegg20}. In summary, observational constraints on the main nitrogen reservoir \textrm{in molecular clouds and how they vary with the evolution from clouds to comets} are still lacking.


\textrm{The molecular D/H ratios of ice can probe the formation history of molecules. As deuterium fractionation is more efficient in the dense, cold environment \citep[e.g., molecular cloud cores;][]{Millar89}, information about the formation stage of molecules is imprinted into molecular D/H ratios.}
\textrm{Recently, \citet{Furuya18} proposed a new approach to constrain the main nitrogen reservoir in \textrm{molecular} clouds using D/H ratios of \ammonia ice, i.e., \dammonia/\ammonia\footnote{Strictly speaking, D/H ratio of \ammonia is 1/3 of \dammonia/\ammonia, since the chance for ammonia during its formation to have D is three times higher than e.g. C$_2$H. But we use the term D/H ratio for the ratio of monodeuterated to normal isotopologues in the text for brevity. We consider this statistical issue when we compare the D/H ratio among molecules in Section \ref{subsec:comparison}.}. They performed gas-ice astrochemical simulations to investingate the evolution of nitrogen-bearing species from the formation of molecular clouds, dense prestellar core, and then to the protostellar stage. They showed that if most nitrogen is already locked up in \ammonia ice in the molecular cloud stage, the \dammonia/\ammonia in bulk ice mantle and thus the ratio in the sublimates in the central warm region of the protostellar envelopes (typically $\lesssim$100\,au and $\gtrsim$100\,K; so-called ``hot corino'') are as low as $\sim4\times 10^{-3}$ similar to or slightly higher than the HDO/H$_2$O ratio. If N atoms are the dominant nitrogen reservoir in the molecular cloud and \ammonia ice formation starts only in the prestellar stage, on the other hand, the \dammonia/\ammonia ratio in the protostellar stage could be a few percent or even higher \citep[e.g.,][]{Furuya18, Aikawa12}. The \dammonia/\ammonia ratio in the hot corinos, where ices sublimate, tells us when the \ammonia ice is mainly formed.}


\textrm{The deuteration of \ammonia in the cold outer envelope of Class 0 protostars has been measured by single-dish observations. The \dammonia/\ammonia ratios in the gas-phase are derived to be as high as a few to a few tens of percent \citep[e.g.,][]{Shah01, Hatchell03}. Even doubly- and triply-deuterated ammonia (NHD$_2$ and ND$_3$) have been detected \citep[e.g.,][]{Loinard01, vanderTak02}. On the other hand, deuteration of \ammonia in the hot corino has not been measured so far, as high resolution and sensitivity observations by interferometers are needed.}


One of the few sources where \ammonia emission has been detected in the hot corino is NGC1333 IRAS4A (hereafter IRAS4A), a deeply embedded low-mass Class 0 protobinary system located at a distance of $\sim$300\,pc \citep{Ortiz-Leon18, Zucker20}. 
IRAS4A harbors two protostars IRAS4A1 and IRAS4A2 \citep[hereafter 4A1 and 4A2,][]{Lay95, Looney00}. The $(J, K) = (2,2)$ and $(3,3)$ inversion transitions of \ammonia at 24\,GHz have been observed the Karl G. Jansky Very Large Array (VLA) toward 4A1 and 4A2 by \citet{Choi07, Choi10, Choi11}. Both 4A1 and 4A2 show compact emission of those transitions, indicative of the \ammonia ice sublimation. \textrm{Most recently, \citet{DeSimone2022} also observed $(J,K)=(3,3)$ to (7,7) inversion transitions of \ammonia with the VLA, revealing abundant warm \ammonia gas sublimated from ice in the hot corino.}

In this paper, we report the first detection of singly-deuterated ammonia (\dammonia) emission in the hot corinos of 4A1 and 4A2, using interferometric observations at centimeter wavelengths with the VLA. In Section \ref{sec:observation}, our observations and data reduction are described. We analyzed the \ammonia and \dammonia emission line profiles to constrain the \dammonia/\ammonia ratios in the central region of 4A1 and 4A2 as described in Section \ref{sec:analysis}. In Section \ref{sec:results}, constraints on the column densities of \ammonia and \dammonia, excitation temperatures, and \dammonia/\ammonia ratios are shown. We discuss the implications from the derived \dammonia/\ammonia ratios and constraints on the formation stage of \ammonia molecules in Section \ref{sec:discussion}. Finally, we summarize our results in Section \ref{sec:summary}.


\section{Observation} \label{sec:observation}
The IRAS4A system was observed with VLA during the Semester 2018B (PI: Magnus V. Persson, project code: 18B-125). Two tracks of observations were performed in C-configuration using $K$-band receivers (18--26.5\,GHz) on 2018 November 26 and 2019 January 19, both centered on 4A2 ($\alpha\mathrm{(J2000)}=03^\mathrm{h}29^\mathrm{m}10\fs440$, $\delta\mathrm{(J2000)}=+031\arcdeg13\arcmin32\farcs160$). The phase, flux, and bandpass calibrators are J0336+3218, 0542+498 (3C147), and J3019+4130, respectively. The absolute flux calibration uncertainty is 10-15\% for $K$-band\footnote{\url{https://science.nrao.edu/facilities/vla/docs/manuals/oss/performance/fdscale}}. 

The observed data were split into 23 spectral windows, 10 of which were dedicated to continuum observations and 13 of which are used to observe the line emission. The spectral resolution and bandwidth of the line spectral windows are 7.812\,kHz ($\sim$0.1\,km\,s$^{-1}$) and 4 MHz ($\sim$50\,km\,s$^{-1}$), respectively. The spectral setup targeted several \ammonia and \dammonia transitions listed in Table \ref{tab:observed_transitions} using seven out of 13 line spectral windows. Other spectral windows targeted $^{15}$\ammonia transitions and \water maser emission, which are not presented in this paper. 
We note that \textrm{our correlator setup cover all the hyperfine satellites for \ammonia (1,1) transition, while only the inner satellites are covered for \ammonia (3,3) and \dammonia 3$_{1,3}$--3$_{0,3}$ and 4$_{1,4}$--4$_{0,4}$ transitions. For \ammonia (4,4) and (5,5) transitions, the correlator setup do not cover the hyperfine satellites and only main lines are covered.}

The data were calibrated using the VLA pipeline in Common Astronomy Software Applications (CASA) 5.4.1 \citep{CASA}. Bad data were identified from an initial pipeline run. These data were flagged and the pipeline was re-run. To protect the spectral lines from RFI flagging and down weighting, we manually specified continuum regions in a \texttt{cont.dat} file for the spectral line windows.
In order to identify the line-free channels in the spectral windows with many lines (e.g., \ammonia (1,1) hyperfines), we first Fourier-transformed the visibilities of line spectral windows without any continuum subtraction. The line-free channels are manually selected by carefully inspecting the image cubes. Afterwards, continuum subtraction is done in the $uv$ plane using \texttt{uvcontsub} task for all line spectral windows specifying the line-free channels selected in the image plane. 

The visibilities in all continuum spectral windows and line-free channels in line spectral windows are averaged to $\sim$4\,MHz channel widths, and deconvolved down to $3\,\times$ root mean square (RMS) noise level of the dirty image using the multi-scale CLEAN algorithm implemented in \texttt{tclean} task with scales of [0,10,30] pixels (with a pixel scale of 0\farcs1) and a Briggs parameter of 0.5. To determine the CLEAN components, we used the \textit{automasking} feature implemented in \texttt{tclean} task with the parameters of $\textit{sidelobethreshold}=2.0$, $\textit{noisethreshold}=4.25$, and $\textit{lownoisethreshold}=1.5$. We manually checked that all the emission components are masked by carefully inspecting the resulting image.

\begin{figure}
\epsscale{1.2}
\plotone{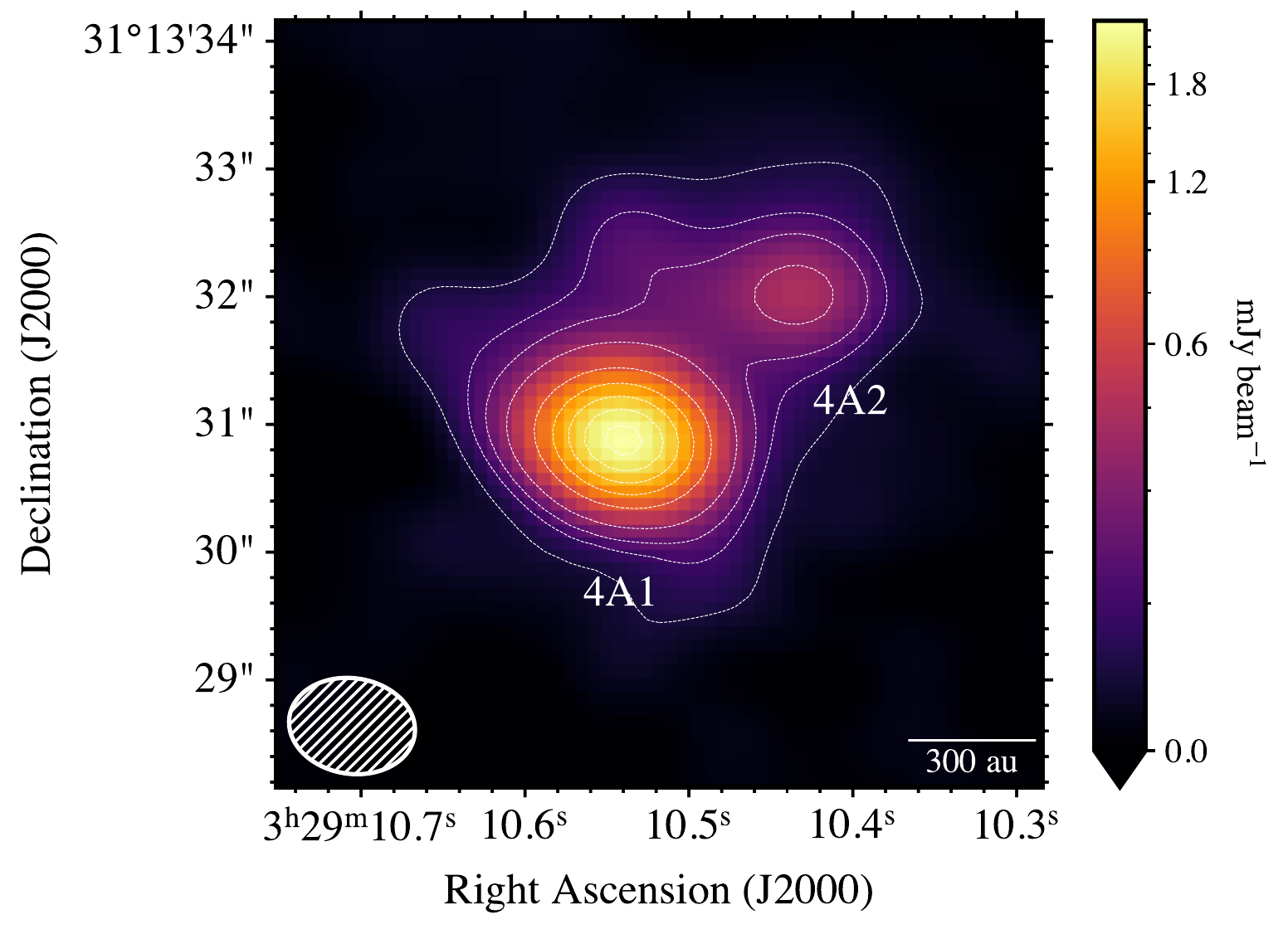}
\caption{The 22.3\,GHz continuum emission maps toward IRAS 4A and IRAS 4B. The contours (white dashed lines) correspond to [10, 20, 30, 50, 100, 150, 200, 250, 300] $\times\,\sigma_\mathrm{RMS}$, where $\sigma_\mathrm{RMS}=7.4$\,$\mu$Jy\,beam$^{-1}$ for both maps. The beam sizes of 0\farcs99 $\times$ 0\farcs75 (P.A. $=$ 82\arcdeg) and scale bars of 300\,au are shown at lower left and right. We note that the color scales employ arcsinh stretches, with the lower end saturating at 0.0.}
\label{fig:continuum}
\end{figure}

Figure \ref{fig:continuum} presents the 22.3\,GHz ($\sim$1.3\,cm) continuum image. The resulting beam size and RMS noise level of the continuum image were 0\farcs99 $\times$ 0\farcs75 (P.A. $=$ 82\arcdeg) and 7.4 $\mu$Jy beam$^{-1}$, respectively. The RMS noise level was measured in the emission-free region. The continuum peak positions are estimated by the 2D Gaussian fitting to be $\alpha\mathrm{(J2000)}=03^\mathrm{h}29^\mathrm{m}10\fs540$, $\delta\mathrm{(J2000)}=+031\arcdeg13\arcmin30\farcs893$ for 4A1 and $\alpha\mathrm{(J2000)}=03^\mathrm{h}29^\mathrm{m}10\fs442$, $\delta\mathrm{(J2000)}=+031\arcdeg13\arcmin32\farcs011$ for 4A2, which are consistent with those derived by \citet{DeSimone20}. The peak intensities of $2.2\pm0.3$\,mJy\,beam$^{-1}$ and $0.42\pm0.06$\,mJy\,beam$^{-1}$ for 4A1 and 4A2, respectively. The flux calibration uncertainty was added in quadrature. These values are consistent with those derived by \citet{DeSimone20} within uncertainty.

The visibilities in line spectral windows are imaged using the multi-scale CLEAN algorithm with scales of [0, 10, 30] pixels, a Briggs parameter of 2.0, 1.0\,km\,s$^{-1}$ velocity channel widths, and the same \textit{automasking} parameters as the continuum imaging. 
The properties of images such as beam sizes and RMS noise levels ($\sigma_\mathrm{RMS}$) are listed in Table \ref{tab:image_property}. 

\begin{figure*}
\epsscale{1.18}
\plotone{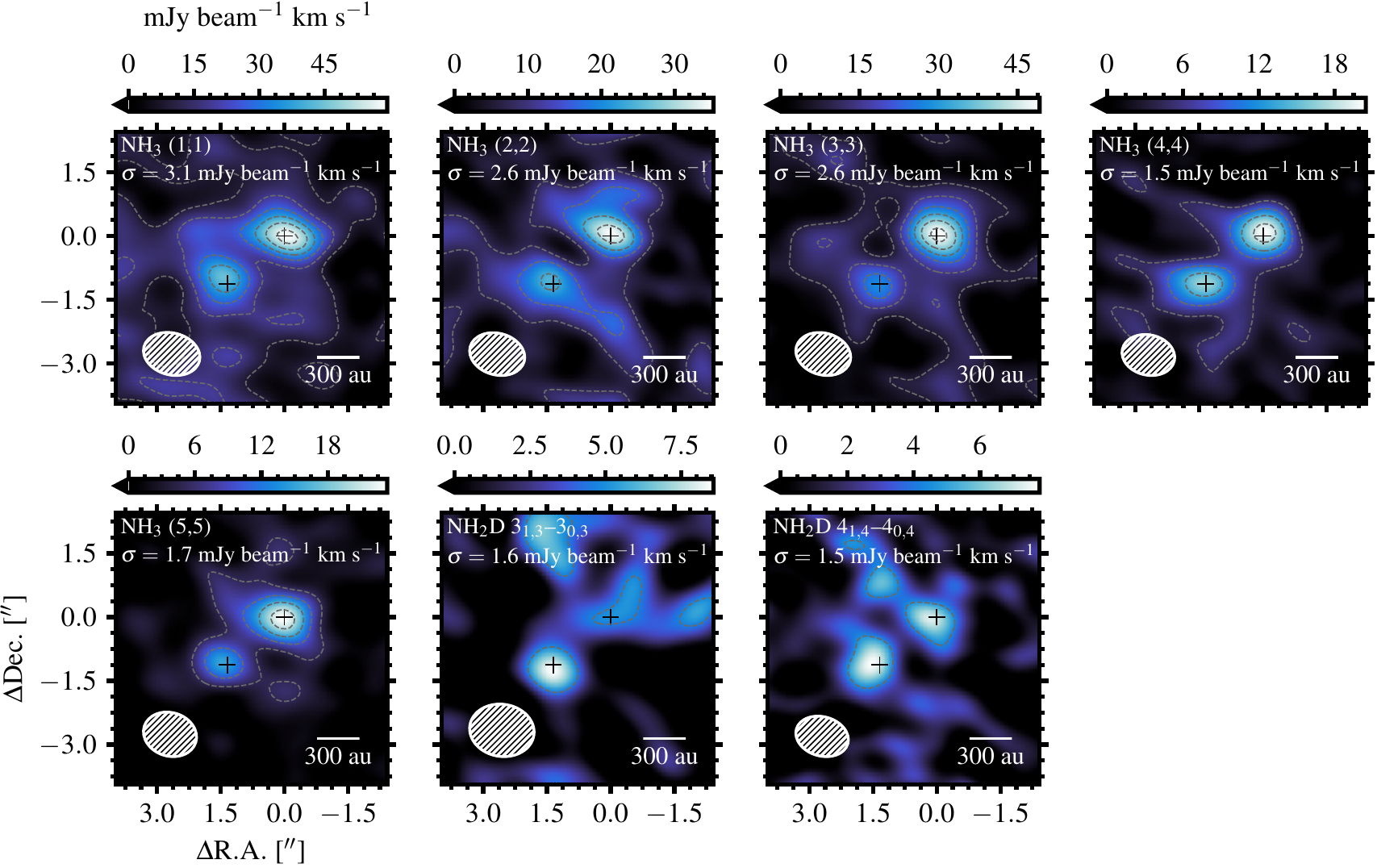}
\caption{Gallery of velocity-integrated intensity maps of \ammonia and \dammonia transitions. Transitions are indicated at the upper left in each panel. The continuum emission peaks for 4A1 and 4A2 are indicated by black crosses in each panel. \textrm{The grey dashed contours start from 3$\sigma$, followed by steps of 3$\sigma$.} The beam size and scale bar of 300\,au are shown in lower right and left in each panel, respectively. We note that the color scale is saturated at the lower end of 0.0. }
\label{fig:mom0_gallery}
\end{figure*}

Figure \ref{fig:mom0_gallery} shows the gallery of velocity-integrated emission maps for all \ammonia and \dammonia lines, generated using \texttt{bettermoments} \citep{Teague18_bettermoments}. The image cubes are integrated over the velocity ranges listed in Table \ref{tab:image_property} encompassing all detected hyperfine components. \textrm{The velocity ranges for integration are adjusted for each transition to include only the emission by visually inspecting the image cubes.} The noise levels $\sigma$ for these maps are calculated by \texttt{bettermoments} as $\sigma = \sigma_\mathrm{RMS}\times\sqrt{N}\times dv$, where $N$ is the number of integrated channels and $dv$ is the channel width ($1.0$\,km\,s$^{-1}$). These values are reported in each panel of Figure \ref{fig:mom0_gallery}. We clearly see the detection of all  \ammonia lines toward both 4A1 and 4A2 at S/Ns of $\sim$10--20, while \dammonia lines are detected toward 4A1 with S/N of $\sim$5 and only tentatively detected toward 4A2 with S/Ns of $\sim$3--\textrm{5}. \textrm{These S/N values are reported in Table \ref{tab:image_property} as well.}



\begin{deluxetable*}{cCCCCR}
\tablecaption{Observed Transitions and Their Spectroscopic Data}
\label{tab:observed_transitions}
\tablehead{\colhead{Specie} & \colhead{Transition} & \colhead{$\nu_0$} & \colhead{$\log_{10} A_\mathrm{ul}$} & \colhead{$g_\mathrm{u}$} & \colhead{$E_\mathrm{u}$} \\
& & \colhead{[GHz]} & \colhead{[s$^{-1}$]} & & \colhead{[K]}
}
\startdata 
\ammonia & (1,1) & 23.6944955 & -6.78260 & 6.0 & 23.26484 \\
& (2,2) & 23.7226333 & -6.65631 & 10.0 & 64.44806 \\
& (3,3) & 23.8701292 & -6.59744 & 28.0 & 123.53904 \\
& (4,4) & 24.1394163 & -6.55545 & 18.0 & 200.52091 \\
& (5,5) & 24.5329887 & -6.51738 & 22.0 & 295.37076 \\
\dammonia & 3_{1,3}-3_{0,3} & 18.807888 & -7.42815 & 63.0 & 93.91931 \\
& 4_{1,4}-4_{0,4} & 25.023792 & -7.01363 & 27.0 & 152.25741\\
\enddata
\tablecomments{\textrm{The spectroscopic data are retrieved from the Jet Propulsion Laboratory (JPL) Catalogue \citep{Pickett1988} and Cologne Database for Molecular Spectroscopy \citep[CDMS,][]{CDMS} via the Splatalogue interface. The original data are presented in \citet{Yu2010} for \ammonia and \citet{DeLucia75,Cohen82,Fusina88} for \dammonia.}}
\end{deluxetable*}

\begin{deluxetable*}{cccCccc}
\footnotesize
\tablecaption{Properties of Image Cubes and Velocity-integrated Intensity Maps of \ammonia and \dammonia}
\label{tab:image_property}
\tablehead{\colhead{Specie} & \colhead{Transition} & \colhead{RMS} & \colhead{Synthesized Beam (P.A.)} & \colhead{Integration Range$^\dagger$} & \multicolumn{2}{c}{\textrm{S/N}$^\ddagger$} \\ 
\cline{6-7}
& & [mJy\,beam$^{-1}$] & & \colhead{[km\,s$^{-1}$]} & \textrm{4A1} & \textrm{4A2}
}
\startdata 
NH$_3$ & $(1,1)$ & 0.69 & 1\farcs35 \times 0\farcs99\,(75\arcdeg) & [$-14$, $-11$], [$-3$, $1$], [$4$, $9$], [$12$, $16$], [$23$, $29$] & \textrm{12} & \textrm{16} \\
NH$_3$ & $(2,2)$ & 0.62 & 1\farcs32 \times 0\farcs98\,(75\arcdeg) & [$-13$, $-8$], [$4$, $10$], [$21$, $26$] & \textrm{9.2} & \textrm{13} \\
NH$_3$ & $(3,3)$ & 0.60 & 1\farcs31 \times 0\farcs98\,(75\arcdeg) & [$-16$, $-11$], [$3$, $11$], [$25$, $30$] & \textrm{10} & \textrm{18} \\
NH$_3$ & $(4,4)$ & 0.55 & 1\farcs25 \times 0\farcs95\,(77\arcdeg) & [$3$, $10$] & \textrm{11} & \textrm{13} \\
NH$_3$ & $(5,5)$ & 0.60 & 1\farcs27 \times 1\farcs02\,(73\arcdeg) & [$3$, $10$] & \textrm{9.0} & \textrm{14} \\
NH$_2$D & 3$_{1,3}$--3$_{0,3}$ & 0.59 & 1\farcs52 \times 1\farcs23\,(81\arcdeg) & [\textrm{4}, $10$] & \textrm{5.5} & \textrm{3.1} \\
NH$_2$D & 4$_{1,4}$--4$_{0,4}$ & 0.58 & 1\farcs26 \times 0\farcs92\,(74\arcdeg) & [$4$, $10$] & \textrm{5.2} & \textrm{4.9} \\
\enddata
\tablenotetext{}{$\dagger$ Integration ranges for velocity-integrated emission maps.}
\tablenotetext{}{$\ddagger$ \textrm{Peak signal-to-noise ratios on the velocity-integrated emission maps.}}
\end{deluxetable*}

\section{Analysis} \label{sec:analysis}
In order to estimate the column densities of \ammonia and \dammonia, first we extracted the spectra for all lines from a single pixel toward the continuum peak position estimated in Section \ref{sec:observation} for each source. The extracted spectra are presented in Figure \ref{fig:spectrum_fit_4A1} and \ref{fig:spectrum_fit_4A2} for 4A1 and 4A2. For the following analysis, \ammonia (1,1) and (2,2) transitions are excluded; given their relatively low excitation energies, the extended emission may contaminate the emission from the central hot region. 


\subsection{Estimates on the emitting region size}
\textrm{To estimate the emitting region size, we inspected the visibility profiles in $(u,v)$ plane (amplitude--baseline length plot) for each transition. However, all of the transitions show almost flat profiles, indicating that it is difficult to infer the emitting region size even from visibility data. Instead, w}e estimated the emitting region size of \ammonia using our data and the data presented in \citet{Choi10, Choi11}. \citet{Choi10, Choi11} observed \ammonia (3,3) transitions toward the IRAS4A system with a circular beam of 0\farcs3 $\times$ 0\farcs3, which is higher spatial resolution than our data. The difference in the beam sizes between the data in \citet{Choi10, Choi11} and our data allows us to evaluate the beam dilution and accurately estimate the emitting region size, assuming that the both observations trace the region with the same physical condition and that the lines are optically thick. 
We will confirm that the main component of \ammonia (3,3) is highly optically thick ($\tau\gtrsim$10) in the following section.

We fitted a Gaussian to the spectra presented in Figure 4 in \citet{Choi11} and Figure 1 in \citet{Choi10} for IRAS4A1 and IRAS4A2, resulting in the peak brightness temperatures of $44.4\pm2.3$\,K and $73.2\pm2.5$\, K, respectively. We also fitted a Gaussian to the \ammonia (3,3) main component of our data to obtain the peak brightness temperature. 
The observed peak brightness temperatures can be modeled as $T_\mathrm{int}\times f$, where $T_\mathrm{int}$ is the intrinsic temperature and $f$ is the beam filling factor given as 
\begin{equation}\label{eq:bff}
    f = \frac{\theta_\mathrm{s}}{\sqrt{\theta_\mathrm{s}^2 + \theta_\mathrm{maj}^2}}\frac{\theta_\mathrm{s}}{\sqrt{\theta_\mathrm{s}^2 + \theta_\mathrm{min}^2}},
\end{equation}
where $\theta_\mathrm{s}$ is the emitting region size, and $\theta_\mathrm{maj}$ and $\theta_\mathrm{min}$ are the size of the beam major and minor axes, respectively. Here we assume that the emission distribution of the source is a symmetric 2D Gaussian with a FWHM of $\theta_\mathrm{s}$. Assuming that $T_\mathrm{int}$ is the same between the data, the relation between the observed peak brightness temperature ratio and $\theta_\mathrm{s}$ can be inferred. We constructed the likelihood function of the observed brightness temperature ratio and $\theta_\mathrm{s}$. To sample the posterior distribution, we employed the Markov Chain Monte Carlo (MCMC) method implemented in \texttt{emcee} python package \citep{emcee}. We used an uniform prior in the interval [0\farcs01,0\farcs6] for $\theta_\mathrm{s}$ and run the 200 walkers for 600 steps, including initial 100 steps discarded as burn in. We found $\theta_\mathrm{s} = 0\farcs25^{+0.05}_{-0.06}$ and $0\farcs30^{+0.03}_{-0.03}$ for 4A1 and 4A2, respectively, where the estimated values are the median of the posterior distributions and uncertainties are the 16th and 84th percentile. 
These estimates are used in the subsequent analyses. The estimated emitting region is close to the beam size of \citet{Choi10}, in which the deconvolved sizes of the sources are derived to be $0\farcs44\times0\farcs25$ for 4A1 and $0\farcs55\times0\farcs30$ for 4A2. These values are slightly larger than our estimates, most probably because they averaged the images of (2,2) and (3,3), while our estimates are based on the (3,3) transition only.

\subsection{Hyperfine fit of \texorpdfstring{\ammonia}{} and \texorpdfstring{\dammonia}{} lines}
We fitted the observed spectra with a synthetic model considering hyperfine splitting \citep{Rosolowsky08}. The details of the model used in the present work are described in Appendix \ref{appendix:HFS_model}. Among the observed \ammonia and \dammonia transitions, both ortho (\ammonia (3,3) and \dammonia (3,3)) and para (\ammonia (4,4), (5,5), and \dammonia (4,4)) transitions are included, which in principle allows us to estimate the ortho-to-para ratio of \ammonia and \dammonia. In our model, we assume local thermo-dynamical equilibrium (LTE) so that the observed spectra can be fitted by a common excitation temperature for \ammonia and \dammonia transitions under the assumption that they originate from the same gas. The LTE assumption should be valid for targeted sources because the H$_2$ density in the vicinity of the protostar ($\lesssim$100\,au) should be $\gtrsim10^6$\,cm$^{-3}$ \citep[e.g.][]{Persson16, DeSimone20}, which is far higher than the critical densities of \ammonia transitions \citep[$\sim$10$^3\,\mathrm{cm}^{-3}$; e.g.,][]{Shirley15}. 

First, we fitted these models to the \ammonia and \dammonia line profiles simultaneously; the ortho-to-para ratios of \ammonia and \dammonia are fixed to one and three, respectively, which are the statistical values expected in the warm ($\gtrsim$100\,K) environments. This resulted in 14 free parameters (Table \ref{tab:fit_param}): velocity dispersions ($\Delta V$) and systemic velocities ($v_\mathrm{sys}$) for all five transitions, the logarithms of column densities of \ammonia and \dammonia ($\log_{10}N(\mathrm{NH_3})$ and $\log_{10}N(\mathrm{NH_2D})$), and the common excitation temperature ($T_\mathrm{ex}$) and emitting region size ($\theta_\mathrm{s}$) among all five transitions. We also ran a fit with the ortho-to-para ratio of \ammonia as a free parameter to check if the assumptions of statistical equilibrium affects derived parameters. We sampled the posterior distributions with MCMC method implemented in \texttt{emcee} Python package. We used uniform priors for all parameters except for the emitting region size. We used tight Gaussian priors for emitting region sizes based on the estimates in previous subsections. We run 200 walkers for 5000 steps, including the initial 2500 steps discarded as burn in. During the fit, we initially generated model spectra with much higher velocity samplings and then resampled down to the ones in the actual data (1.0\,km\,s$^{-1}$).

\begin{deluxetable}{ccc}
\footnotesize
\tablecaption{Free Parameters of Hyperfine Fits for 4A1 and 4A2}
\label{tab:fit_param}
\tablehead{\colhead{Parameter} & \colhead{Prior for 4A1 fit }\tablenotemark{a} & \colhead{Prior for 4A2 fit}\tablenotemark{a}
}
\startdata 
$\Delta V$ [km\,s$^{-1}$]$^\dagger$ & U, [0.1, 10] & U, [0.1, 10] \\
$v_\mathrm{sys}$ [km\,s$^{-1}$]$^\dagger$ & U, [4, 10] & U, [4, 10] \\
$\log_{10}N(\mathrm{NH_3})$ [cm$^{-2}$]& U, [14, 21] & U, [14, 21] \\
$\log_{10}N(\mathrm{NH_2D})$ [cm$^{-2}$] & U, [14, 21] & U, [14, 21] \\
$T_\mathrm{ex}$ [K] & U, [50, 350] & U, [50, 350] \\
$\theta_\mathrm{s}$ [\arcsec] & G, [0.25, 0.06] & G, [0.30, 0.03]\\
o/p$^\ddagger$ & U [0.01, 100] & U [0.01, 100]\\
\enddata
\tablenotetext{a}{The prior distributions of the parameter. The type of prior (uniform or Gaussian) and its parameters are listed. ``U'' represents an uniform prior, followed by the lower and upper bound of the uniform prior. ``G'' represents a Gaussian prior, followed by the center and standard deviation of the Gaussian prior.}
\tablenotetext{}{$\dagger$ The priors are common for all transitions, while treated as individual parameters for each transition.}
\tablenotetext{}{$\ddagger$ The ortho-to-para ratio of \ammonia. Used for fits with free ortho-to-para ratio only.}
\end{deluxetable}

\section{Results}\label{sec:results}

\begin{figure*}
\epsscale{1.15}
\plotone{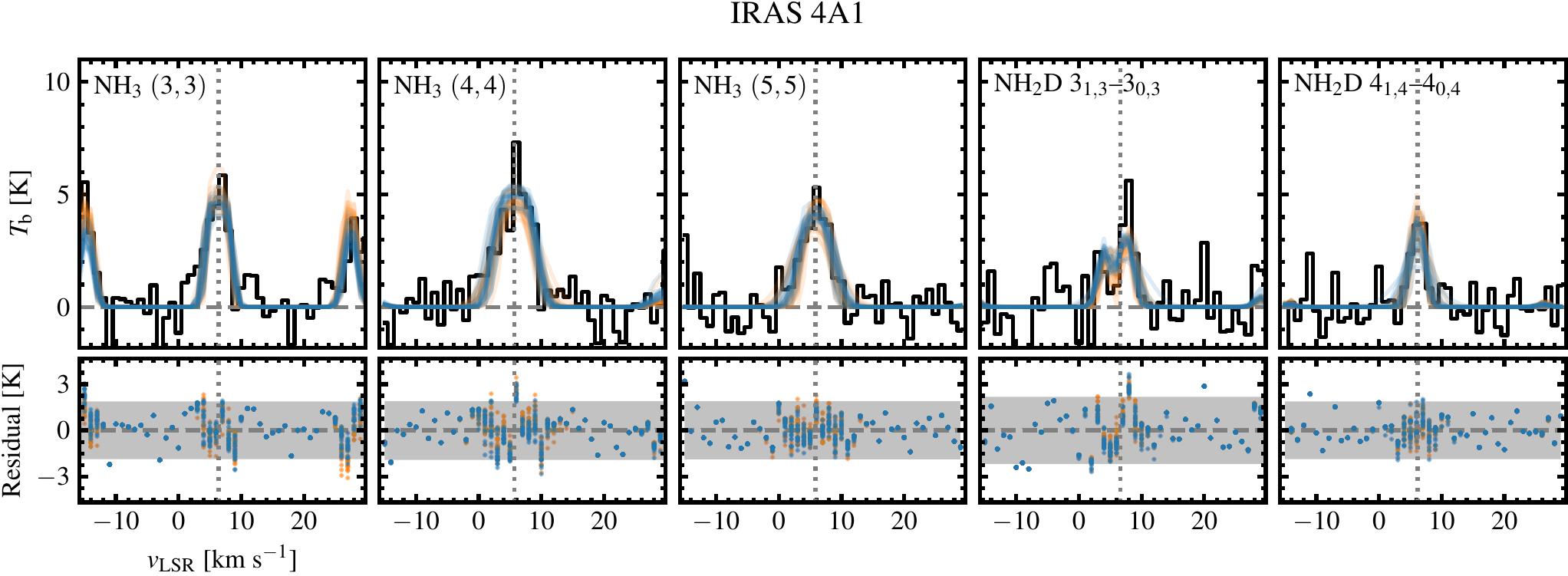}
\caption{\textrm{(top) Observed spectra of \ammonia (3,3), \ammonia (4,4), \ammonia (5,5), \dammonia $3_{1,3}$--$3_{0,3}$, and \dammonia $4_{1,4}$--$4_{1,4}$ (black) extracted toward the continuum peaks of 4A1. The model spectra generated by 20 randomly selected parameter samples from the MCMC chains of the fits with fixed (blue) and free (orange) ortho-to-para ratios of \ammonia are overlaid. (bottom) Residual of observations and fitted model with fixed (blue) and free (orange) ortho-to-para ratio of \ammonia. Each data point correspond to the randomly selected models from MCMC chains. The grey-shaded regions indicate 2$\sigma_\mathrm{RMS}$ range. For both top and bottom panels, while the zero flux levels are indicated by grey dashed horizontal lines, the vertical dotted line indicate the systemic velocity.}}
\label{fig:spectrum_fit_4A1}
\end{figure*}

\begin{figure*}
\epsscale{1.15}
\plotone{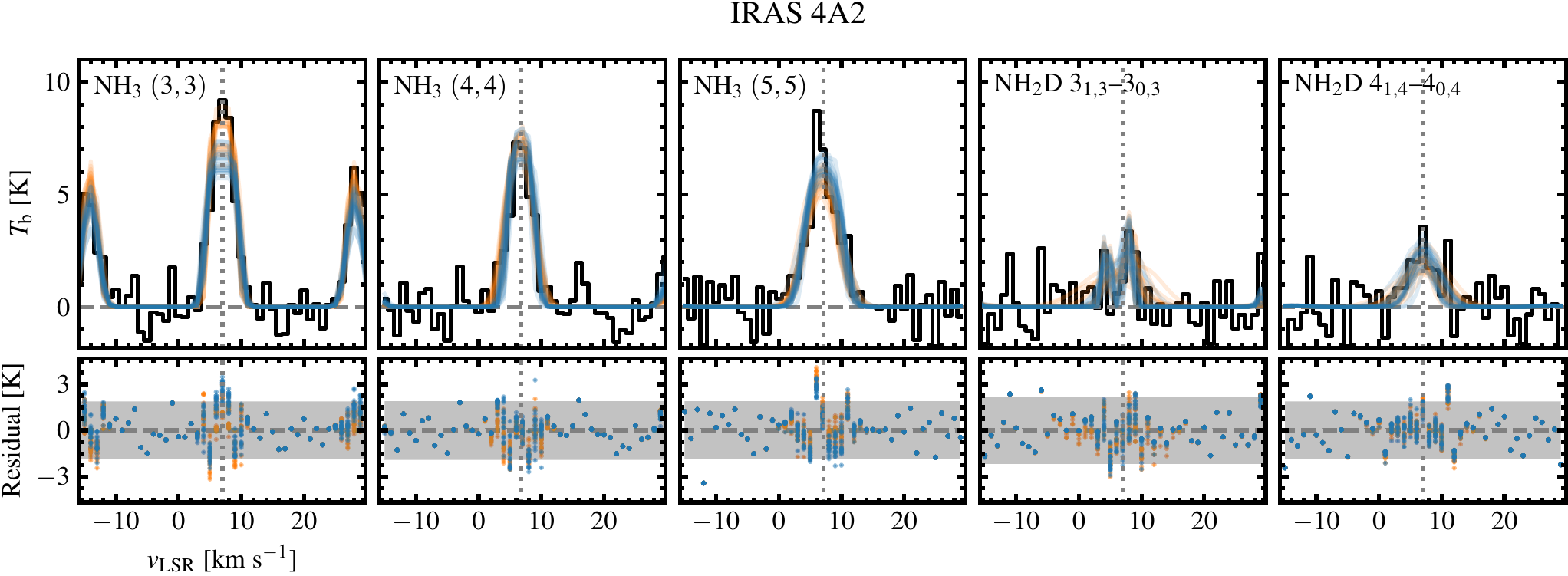}
\caption{\textrm{Same as Figure \ref{fig:spectrum_fit_4A1}, but for 4A2.}}
\label{fig:spectrum_fit_4A2}
\end{figure*}

The observed spectra overlaid by the model spectra generated from posterior samples\textrm{, as well as the residual spectra after subtracting the model spectra} are presented in Figure \ref{fig:spectrum_fit_4A1} and \ref{fig:spectrum_fit_4A2}. The results of the fits are summarized in Table \ref{tab:fit_results}.

For 4A1, high excitation temperatures are derived in both fits with ortho-to-para ratio fixed and varied (98$_{-7}^{+8}$\,K and 110$_{-10}^{+10}$\, K, respectively). The total optical depth (including the hyperfine) of each \ammonia transition is greater than unity, while the hyperfine satellites of the \ammonia (3,3) transition provide the constraints on the column density of \ammonia. In the fits with free ortho-to-para ratio, the \ammonia column density is more uncertain and the \ammonia ortho-to-para ratio is not well constrained. Since a statistical \ammonia ortho-to-para ratio of unity is expected in the hot region, we employ the fit with the fixed ortho-to-para ratio as the fiducial result. The \dammonia column densities result in similar values ($\sim7\times10^{17}$\,cm$^{-2}$) in both fits. Particularly, the undetected hyperfine satellites of \dammonia (4$_{1,4}$--4$_{0,4}$) at $\sim-16$\,km\,s$^{-1}$ and $\sim$26\,km\,s$^{-1}$ (see Figure \ref{fig:spectrum_fit_4A1}) are helpful to constrain the column density. The line widths of \ammonia (4,4) and (5,5) transitions are broader than those of \ammonia (3,3) and \dammonia transitions (Table \ref{tab:fit_results}). This may indicate that \textrm{these higher transitions of \ammonia preferentially trace inner regions with higher infall velocity than (3,3) transition of \ammonia and \dammonia transitions. We will discuss this issue in more detail in Section \ref{sec:discussion}}
Finally, both fits yield high \dammonia/\ammonia column density ratios of \textrm{$0.96_{-0.39}^{+0.76}$ and $0.28_{-0.23}^{+0.59}$}, consistent within the uncertainties. 

For 4A2, the fit with a fixed ortho-to-para ratio yields a lower excitation temperature ($87_{-10}^{+13}$\,K) than that with a free ortho-to-para ratio (\textrm{$160_{-20}^{+30}$}\,K). In the former, the peak brightness temperature of the main component of \ammonia (3,3) transition is significantly underestimated (Figure \ref{fig:spectrum_fit_4A2}) and thus the lower excitation temperature is not reasonable. In contrast, the spectra are better reproduced in the fit with a free ortho-to-para ratio, which suggests that the higher excitation temperature is more reasonable. However, the obtained ortho-to-para ratio of \ammonia in the fit (\textrm{$4.5_{-1.3}^{+1.5}$}) is significantly higher than the statistical ratio of unity. This does not seem to be realistic since the ortho-to-para ratio of \ammonia should be unity in the hot region as indicated by the derived excitation temperature. 
Nevertheless, the column density estimates in the two fits agreed within the uncertainties. We thus adopt the \ammonia column density derived from the fits with free ortho-to-para ratio as our fiducial value. The \dammonia column density is also well constrained, although the line widths are not well constrained due to the lower S/N. The widths of \ammonia lines are typically $\sim$2--5\,km\,s$^{-1}$, and broader for higher excitation transitions (Table \ref{tab:fit_results}). This could again imply that the emitting region is smaller (i.e., inner region with higher velocity) for higher transitions. Finally, the fiducial fit yield a \dammonia/\ammonia ratio of \textrm{$0.48_{-0.13}^{+0.17}$}, consistent with that derived in the fit with fixed ortho-to-para ratio within the uncertainties.



\movetabledown=50mm
\begin{rotatetable*}
\begin{deluxetable*}{cCCCCCCCCCCC}
\tablecaption{Results of Fits for 4A1 and 4A2}
\label{tab:fit_results}
\centering
\tablehead{\colhead{o/p$^\dagger$} & \colhead{$\theta_\mathrm{s}$ [\arcsec]} & \colhead{$T_\mathrm{ex}$ [K]} & \multicolumn{5}{c}{$\Delta V_\mathrm{FWHM}$ [km\,s$^{-1}$]} & \colhead{$N$(\ammonia) [cm$^{-2}$]} & \colhead{$N$(\dammonia) [cm$^{-2}$]} & \colhead{\dammonia/\ammonia} & \colhead{Fiducial} \\
\cline{4-8}
& & & \colhead{\ammonia (3,3)} & \colhead{\ammonia (4,4)} & \colhead{\ammonia (5,5)} & \colhead{\dammonia 3$_{1,3}$--3$_{0,3}$} & \colhead{\dammonia\,4$_{1,4}$--4$_{0,4}$} & & & & 
}
\startdata 
\multicolumn{12}{c}{IRAS 4A1} \\ \hline 
1 (fixed) & 0.25_{-0.06}^{+0.06} & 98_{-7}^{+8} & 1.9_{-0.2}^{+0.3} & 3.9_{-0.6}^{+0.7} & 3.5_{-0.7}^{+0.9} & 1.8_{-0.5}^{+0.8} & 1.6_{-0.8}^{+1.1} & 8.1_{-2.5}^{+4.0} \times 10^{17} & 7.3_{-1.9}^{+5.4} \times 10^{17} & 0.96_{-0.39}^{+0.76} & \checkmark \\
$11_{-8}^{+40}$ & 0.25_{-0.06}^{+0.06} & 110_{-10}^{+10} & 1.3_{-0.2}^{+0.3} & 4.7_{-0.8}^{+1.0} & 4.3_{-0.8}^{+1.0} & 1.9_{-0.5}^{+0.8} & 1.6_{-0.8}^{+1.1} & 2.8_{-2.0}^{+11.0} \times 10^{18} & 6.9_{-1.6}^{+3.8} \times 10^{17} & 0.28_{-0.23}^{+0.59} & \\ 
\hline \multicolumn{12}{c}{IRAS 4A2} \\ \hline 
1 (fixed) & 0.33_{-0.03}^{+0.02} & 87_{-10}^{+13} & 2.2_{-0.2}^{+0.2} & 2.2_{-0.2}^{+0.3} & 4.1_{-0.7}^{+0.7} & 0.9_{-0.5}^{+2.0} & 5.1_{-1.6}^{+2.0} & 7.9_{-1.6}^{+2.5} \times 10^{17} & 3.1_{-1.0}^{+1.2} \times 10^{17} & 0.38_{-0.13}^{+0.17} & \\ 
$4.5_{-1.3}^{+1.5}$ & 0.26_{-0.02}^{+0.02} & 160_{-20}^{+30} & 2.0_{-0.2}^{+0.2} & 3.0_{-0.5}^{+0.6} & 4.7_{-0.6}^{+0.6} & 2.0_{-1.3}^{+5.2} & 5.1_{-1.5}^{+2.0} & 1.0_{-0.2}^{+0.3} \times 10^{18} & 4.9_{-1.4}^{+1.9} \times 10^{17} & 0.48_{-0.13}^{+0.17} & \checkmark\\ 
\enddata
\tablenotetext{}{$^\dagger$ The ortho-to-para ratio of \ammonia}
\end{deluxetable*}
\end{rotatetable*}

\section{Discussion} \label{sec:discussion}

\subsection{\texorpdfstring{\ammonia}{} and \texorpdfstring{\dammonia}{} in the hot corino region}\label{subsec:NH3_NH2D}

We have detected the high excitation transitions of \ammonia (4,4) and (5,5) toward both 4A1 and 4A2 in addition to the (2,2) and (3,3) lines, which were previously detected by \citet{Choi07, Choi10}. The excitation temperatures derived in the hyperfine fits ($\sim$100\,K for 4A1 and $\sim$150\,K for 4A2) are comparable to or higher than the typical \ammonia sublimation temperature of 100--140\,K depending on the gas density and composition of ice mixture \citep[][]{Hama2013,Furuya14,Minissale2022}. Since the gas and dust temperatures are expected to be well coupled in the high density region, the high excitation temperatures indicate that the observed emission traces the \ammonia ice sublimation zone in the vicinity of the protostars. 
This conclusion is also supported by the compact emitting regions ($\sim$0\farcs25 or $\sim$75\,au) and the high column density of \ammonia ($\sim$10$^{17}$--10$^{18}$\,cm$^{-2}$).
\citet{Taquet13} derived the column density of sublimated H$_2$O to be (3--5) $\times$\,$10^{19}$ cm$^{-2}$ towards IRAS 4A, assuming the emitting region size of $\sim0\farcs4$. The NH$_3$/H$_2$O abundance ratio of $10^{-2}$--$10^{-1}$ is consistent with the composition of interstellar ices observed towards low-mass protostars \citep{Oberg11}.
In addition to \ammonia, we have detected the high excitation transitions of \dammonia. The observed line profiles are reasonably reproduced by the same emitting region sizes and excitation temperature as those of \ammonia, suggesting that the observed emission traces \dammonia sublimated from ices. The \dammonia column densities are also high, resulting in high \dammonia/\ammonia ratios of $\sim$0.5--1 toward both 4A1 and 4A2.

The high \dammonia/\ammonia ratios, higher than expected from \citet{Furuya18}, motivate careful discussions and confirmation. Firstly, the \ammonia column densities may be underestimated, as the \ammonia lines are optically thick. Particularly, our spectral windows do not cover the hyperfine satellites of \ammonia (4,4) and (5,5) transitions that can be optically thin ($\tau<1$). \textrm{\citet{DeSimone2022}} observed the higher excitation lines up to $(J,K) = (7,7)$ covering all of the hyperfine satellites, which allows them to accurately estimate the column density of \ammonia with non-LTE LVG analysis. While our estimates of \ammonia column density in 4A2 ($\sim$ 1 $\times$ 10$^{18}$\,cm$^{-2}$) is consistent with the estimates by \textrm{\citet[][(0.6--3) $\times$ 10$^{18}$\,cm$^{-2}$ with a best-fit of 2 $\times$ 10$^{18}$\,cm$^{-2}$]{DeSimone2022}, our estimate for 4A1 ($\sim$8 $\times$ 10$^{17}$\,cm$^{-2}$) is lower by a factor of a few than the estimates by \citet[][(1--5) $\times$ 10$^{18}$\,cm$^{-2}$ with a best-fit of 2 $\times$ 10$^{18}$\,cm$^{-2}$]{DeSimone2022}. In their work, the optically thin (7,7) transition allowed them to constrain the column density. Our analysis may indeed underestimate the \ammonia column density, particularly in 4A1.} \textrm{It is also possible that the column density of para-\ammonia that was constrained from \ammonia (4,4) and (5,5) is underestimated due to the lack of hyperfines; this might cause the peculiar ortho-to-para ratio in one of the fits.} \textrm{\dammonia/\ammonia best-fit ratios are still higher than the model prediction, 0.4 for 4A1 and 0.3 for 4A2, if we adopt the best-fit \ammonia column densities derived by \citet{DeSimone2022}}.

Another possible caveat is that the kinetic gas motion could deviate the line profile from a simple Gaussian, which also could alter the estimation of column density and \dammonia/\ammonia ratios. The gas infall motion of the envelope can result in the blue-skewed or inverse P-Cygni profile, which has already been observed toward IRAS4A system \citep[e.g. CH$_3$OH and H$_2$O lines,][]{DiFrancesco01,Kristensen12, Mottram13,Su19}.  Indeed, the observed spectra of \ammonia (3,3) and (4,4) toward 4A1 in Figure \ref{fig:spectrum_fit_4A1} show subtle deviations from Gaussian with rapid intensity decreases at the redshifted wings, which may trace the infall motion of the envelope, although it could also be due to the noise. The disk rotation could affect the line profile as well. The Keplerian rotation of the circumstellar disk with a significant inclination will result in a double-peaked profile. 
Although more sophisticated modeling that include these physical structures is desirable to more accurately evaluate the deuteration of \ammonia, our results of high deuteration would not change qualitatively as no clear evidence of infall or rotation is detected at the current spatial/spectral resolution and sensitivity.

Finally, we note that the line widths are different among the observed transitions (Table \ref{tab:fit_results}). Particularly, the line widths of \ammonia (4,4) and (5,5) are broad compared to the other transitions\footnote{Although \dammonia 4$_{1,4}$--4$_{0,4}$ toward 4A2 is also broad, we do not consider this as a robust result due to the relatively low S/N of the line.}, which may indicate that these transitions trace the more inner regions where the gas infall velocities are larger. If free-fall ($v \propto r^{-0.5}$) is assumed as the gas kinetics, the difference in line widths between \ammonia (4,4) and (5,5) ($\sim$4\,km\,s$^{-1}$) and the other transitions ($\sim$2\,km\,s$^{-1}$) indicates the 4 times smaller radius of the emitting region for the former.
We thus conducted the fit described in Section \ref{sec:analysis} using only \ammonia (3,3) and \dammonia transitions, all of which show a similar line width. Assuming an ortho-to-para ratio of \ammonia of unity, the column densities of \ammonia and \dammonia, and thus \dammonia/\ammonia ratios are consistent with those of fiducial fits in Table \ref{tab:fit_results} within uncertainties. Therefore, the high \dammonia/\ammonia ratio should be trustworthy at the spatial scales traced by \ammonia (3,3) and \dammonia transitions (i.e., $\sim$0\farcs25 or $\sim$75\,au). 

Yet it is possible that the distributions of \ammonia (traced by (4,4) and (5,5)) and \dammonia are different at smaller spatial scales.
Namely, the narrower line widths of \dammonia may imply that the line emission originates from more outer region than (4, 4) and (5, 5) \ammonia lines, e.g., a ring-like distribution of \dammonia. The \dammonia/\ammonia ratio derived in $\sim 0\farcs25$ scale then would be lower limit for the ring regions and upper limit for the inner region with higher velocity, while we need higher spatial and/or spectral resolution to confirm such radial distribution. The ring region of very high \dammonia/\ammonia ratio could be due to sublimation of multi-layered ice mantle (see Section \ref{subsec:model_comparison})

\subsection{Comparison with theoretical models and origin of high \texorpdfstring{\dammonia/\ammonia}{} ratios}\label{subsec:model_comparison}

\citet{Shah01} detected \dammonia rotation-inversion transitions at 86\,GHz and 110\,GHz with the NRAO 12\,m telescope. They derived \dammonia/\ammonia ratios of $\sim$0.07 at $\sim$24000\,au scale in the circumbinary envelope of IRAS4A system. \citet{Hatchell03} also observed the 86\,GHz transitions with the IRAM 30\,m telescope to obtain \dammonia/\ammonia ratio of $\sim$0.25 at $\sim$7000\,au scale. These values are consistent with predictions of theoretical models \citep[e.g.][]{Aikawa12, Taquet14,Furuya18}. In those models, the molecular D/H ratios in the cold outer envelope (typically at radii outside a few hundreds of au) are controlled by the exothermic exchange reactions in the gas-phase. For example, NH$_3$ can be formed from NH, which is a product of dissociative recombination of N$_2$H$^+$; NH reacts with H$_3^+$ to form NH$_2^+$, which is converted to NH$_4^+$ via sequential reactions of H$_2$. NH$_3$ is formed by the dissociative recombination of NH$_4^+$. In dense cold regions, H$_3^+$ is highly deuterated due to the exchange reaction and CO depletion, and its high D/H ratio propagates to N$_2$H$^+$ and thus to NH$_3$ \citep[e.g.][]{Aikawa2005}. If N atom is abundant in the cold region, NH$_3$ can be formed more efficiently via hydrogenation of N atoms on grain surfaces than via gas-phase reactions. They are also deuterated by abundant D atoms, which are formed by dissociative recombination of H$_2$D$^+$. NH$_3$ and NH$_2$D formed on cold grain surfaces, however, remain in the ice phase, and desorbed only inefficiently via non-thermal desorption at low temperatures \citep{Hama2013, Martin-Domenech2014, Tinacci2022}.

In the central region (typically $\lesssim$100\,au), on the other hand, temperature increases to $\gtrsim$100\,K. Volatiles such as \ammonia start to sublimate from the grain surface and their gas-phase abundances rapidly increase. In this region, the molecular D/H ratios reflect those of sublimated ice \citep[e.g, ][]{Furuya18}. While the sublimated molecules are subject to gas-phase reactions, the typical destruction timescale via gas-phase reactions is $\sim$ several $10^4$ yr \citep[e.g.][]{nomura2004}. Even if we consider the enhanced ionization by X-rays \citep{Notsu2021}, the infall timescale would be shorter than chemical timescale in the compact hot corino region ($\lesssim 100$ au). The high excitation temperature ($\gtrsim 100$ K) and high column densities of NH$_3$ and NH$_2$D are indeed consistent with the ice sublimation. 
The derived \dammonia/\ammonia ratios \textrm{($\sim$0.5--1 from our analysis, or $\sim$0.3--0.4 adopting the \ammonia column density of \citet{DeSimone2022})} are, however, significantly higher than  predicted in \citet{Furuya18}, i.e. a few \% even in their N atom rich model. 

Since our \dammonia/\ammonia ratios are similar to the ratio obtained in the outer envelope ($\sim 0.25$) by \citet{Hatchell03}, one may wonder if our observations partially trace the emission of \dammonia from the cold or intermediate temperature regions. While the estimated emitting region sizes, which can be used as a proxy of the sublimation region of \ammonia, are $\sim$0\farcs25 or $\sim$75\,au, the spatial resolution of our observations are much lower, i.e. $\sim$1\arcsec or $\sim$300\,au. In addition, even intermediate temperatures (e.g., $\sim$50\,K) could excite observed \ammonia and \dammonia transitions with the energy levels of 90--150\,K from the ground state. We thus conducted a non-LTE radiative transfer simulation of a free-falling envelope model and compared the simulated spectra with the observed spectra (see Appendix \ref{appendix:RT_model} for details). We confirmed that the observed \dammonia emission is dominated by the emission from the hot corinos. 

The very high \dammonia/\ammonia ratio and the possible ring-like distribution of \dammonia could be due to multi-layered structure of ice mantle. \citet{Taquet14}, who calculated gas-grain chemistry of a star-forming core as in \citet{Furuya18}, explicitly showed the fractional abundances of icy molecules in each monolayer of ice mantle. While \dammonia/\ammonia is $\lesssim 10^{-2}$ in deep layers which are formed in the early times, the ratio is $\sim 0.3$ in the surface layers of ice
when the dense prestellar core is about to collapse. As the ice-coated grains enter the central warm region of the core, the highly deuterated ice on the grain surface will be the first to sublimate.


Even though \dammonia may be distributed in a ring region, our observation suggests that N atom is the major nitrogen reservoir in molecular clouds. The \ammonia column density estimated from the (3, 3) transition, which has a similar line width as that of \dammonia, is similar to the column density derived from the combination with higher transitions (see Section \ref{subsec:model_comparison}). The ammonia formed in the cold prestellar stage with a high D/H ratio is thus at least comparable in amount to that in the \textrm{innermost} regions. Since icy ammonia is more efficiently formed via hydrogenation of N atoms on the grain surfaces than in the gas-phase, atomic nitrogen should be abundant and remains as the main nitrogen reservoir in prestellar core stage, where enhanced deuteration is possible.
In the model of \citet{Taquet14}, which predicts relatively abundant \ammonia ice and very high \dammonia/\ammonia in the surface of ice mantle, nitrogen is assumed to be all atomic in their initial condition.

\citet{Furuya18} showed that the [NHD$_2$/NH$_2$D]/[NH$_2$D/NH$_3$] ratio could better trace the primary nitrogen reservoir than the \dammonia/\ammonia ratio alone. If the atomic nitrogen is largely converted into the molecular form such as \Ntwo and \ammonia in the early stage of molecular clouds, most of ammonia ices should have formed by then, which results in relatively low \dammonia/\ammonia ratio. NHD$_2$ is yet formed mainly in the later prestellar stage, since it requires more deuterium atoms, and the NHD$_2$/\dammonia ratio will be higher than \dammonia/\ammonia, i.e., \DDratio $> 1$. On the other hand, if atomic nitrogen remains the dominant nitrogen reservoir in the prestellar phase, which is likely the case for IRAS4A, all the deuterated isotopologues will be formed in the prestellar phase, and \DDratio will be the statistical ratio ($\sim$1/3). 
Thus, observations of doubly-deuterated \ammonia emission are desirable to confirm our conclusion.

\subsection{Comparison with other major species in hot corinos and comet 67P}\label{subsec:comparison}

The first measurement of \dammonia/\ammonia ratios achieved in this work enables us to compare the molecular D/H ratios among molecules in hot corinos. Comparison with cometary D/H ratios is also important, since the molecules in hot corinos could be incorporated to protoplanetary disks and then to planetary material.
The D/H ratios of major species such as water and methanol in the inner warm regions of protostellar cores have extensively been studied toward several sources including IRAS4A \citep[e.g.,][]{Persson14, Jensen19, Taquet19}. Recent in situ measurements on the comet 67P/Churyumov-Gerasimenko (hereafter 67P) by Rosetta project \citep{Altwegg19} has revealed the molecular D/H ratios in major volatiles. The observed molecular D/H ratios toward IRAS4A and comet 67P are summarized in Table \ref{tab:deuteration_summary}. We note that the D/H ratios of water and methanol toward IRAS4A are available only toward 4A2 because the deuterated as well as normal isotopologues have not been detected in emission toward 4A1 in (sub-)millimeter wavelengths probably due to the absorption by the highly optically thick dust. \textrm{We note that the following comparison stands even if we adopt the \ammonia column density derived in \citet{DeSimone2022}, which results in \dammonia/\ammonia $\sim$0.3.} 

Among water, methanol, and ammonia, water shows the lowest D/H ratios of $\sim$10$^{-4}$ toward IRAS4A \citep[][see also \citealt{Jensen19}]{Persson14}. This has been interpreted as evidence that the water ices are mainly formed in the early molecular cloud phase \citep{Furuya16}. We note that in the comparison of the D/H ratio among molecules, we need to take into account the number of hydrogen in the normal isotopologue; e.g., the chance for \ammonia to get one hydrogen replaced by deuterium is 3/2 times higher than for \water. The HDO/\water ratio is still significantly low compared with the deuteration fraction of \methanol and \ammonia. 

In contrast to water, methanol shows higher D/H ratios of $\sim$10$^{-2}$ toward IRAS4A \citep{Taquet19}, suggesting the formation of methanol ices in the cold prestellar core phase. Higher D/H ratios of methanol than water are also seen in other protostellar cores such as IRAS2A and IRAS~16293-2422 \citep{Persson14, Taquet19, Jorgensen18, Manigand20}. The D/H ratio of ammonia measured toward IRAS4A in this work ($\gtrsim$10$^{-1}$) is higher than that of methanol. \textrm{We note that the D/H ratio of \ammonia is still higher than that of methanol, even if we consider the statistical correction and uncertainties associated with the observations. 
The lower limit of \dammonia/\ammonia ratio is $\sim$0.3 in 4A2 if we consider the uncertainty discussed in Section \ref{subsec:NH3_NH2D}. This is higher than the upper limit of the (CH$_2$DOH + CH$_3$OD)/\methanol ratio by a factor of $\sim$6 (Table \ref{tab:deuteration_summary}). With the statistical correction, the D/H ratio of \ammonia is higher than that of \methanol by a factor of $\sim$9.
} 
This indicates that the ammonia ices are mostly formed in the later stage of prestellar cores, \textrm{possibly} even later than the formation of methanol ices, where the deuteration is more efficient. Observations of ammonia deuteration towards other hot corinos are desirable to confirm that the high D/H ratio and late formation of ammonia ices are general.

The D/H ratios of water and methanol in the comet 67P shows a similar trend  to those in IRAS4A \citep[i.e., water D/H is lower than methanol,][]{Altwegg17, Altwegg19, Drozdovskaya19}, suggesting the inheritance of the volatiles in the star-forming core to the planetary materials. The correlation in COMs abundances between the hot corinos and the comets also supports the inheritance \citep{Drozdovskaya19, Bianchi19}. On the other hand, the D/H ratio of ammonia in the comet 67P is lower than methanol, which is different from the relation in the IRAS4A. It suggests some chemical reprocessing for ammonia ices, or nitrogen as a whole, during the accretion onto the protostellar disk or subsequent protoplanetary disk stage. In the coma of comet 67P, \citet{Altwegg20} recently detected a substantial amount of ammonium salts, which implies the chemical processes of ammonia.
Alternatively, 67P inherit the low D/H component of ammonia, which is probed as high-velocity component in our observation, while the high D/H component is fully released to the gas-phase as discussed in Section \ref{subsec:NH3_NH2D}. Indeed, \citet{Furuya17} showed that stellar UV radiation can selectively remove the high D/H component from the surface of the ice via photodesorption, and the D/H ratios of water can be lower in the protoplanetary disk than those in protostellar envelopes. The same mechanism may work for ammonia as well.

\begin{deluxetable*}{ccccc}
\footnotesize
\tablecaption{Molecular D/H ratios measured toward the inner warm region of the IRAS4A2 core and the comet 67P/C-G}
\label{tab:deuteration_summary}
\tablehead{\colhead{Source} & \colhead{HDO/\water} & \colhead{(CH$_2$DOH + CH$_3$OD)/\methanol} & \colhead{\dammonia/\ammonia} & \colhead{References}}
\startdata
IRAS4A & $5.4\pm1.5\times 10^{-4}$ & (1.5--4.7) $\times$\,10$^{-2}$ & $4.8^{+1.7}_{-1.3} \times 10^{-1}$ & 1,2,3,4 \\
67P/C-G & $1.05\pm0.14\times 10^{-3}$ & $5.6\pm1.2\times10^{-2}$ & $\sim1\times10^{-3}$ & 5,6,7 \\
\enddata
\tablecomments{The values in IRAS4A2 are shown as the representatives of the IRAS4A system. 
}
\tablenotetext{}{\textbf{References.} (1) \citet{Persson14}; (2) \citet{Jensen19}; (3) \citet{Taquet19}; (4) this work; (5) \citet{Altwegg17}; (6) \citet{Drozdovskaya21}; (7) \citet{Altwegg19}.}
\end{deluxetable*}

\subsection{Physical properties of 4A1 and 4A2 cores probed by \texorpdfstring{\ammonia}{} and \texorpdfstring{\dammonia}{}}\label{subsec:core_physical_properties}

Although the detailed modeling of the line profiles are difficult due to the relatively low S/N and poor velocity resolutions, several implications about the physical and chemical structures of the 4A1 and 4A2 cores including their similarities and differences can be obtained based on the results of our hyperfine fits. 

The derived excitation temperatures are significantly different between 4A1 and 4A2; i.e. the excitation temperature in 4A2 ($\sim$150\,K) is higher than that in 4A1 ($\sim$100\,K). This difference is also clearly indicated by the higher peak brightness temperature in 4A2 (Figure \ref{fig:spectrum_fit_4A2}). Since the excitation is expected to be thermalized, i.e., satisfies the LTE condition, the excitation temperatures can be directly used as a proxy of kinetic temperature. A similar trend has already been suggested by \citet{DeSimone20}; they have detected the optically thick \methanol emission to estimate the kinetic temperature of 100\,K and 160\,K for 4A1 and 4A2, respectively. The emitting region sizes of \methanol in \citet{DeSimone20} (0\farcs20--0\farcs30) are similar to the \ammonia emitting region sizes derived in the present work, suggesting that both \methanol and \ammonia trace the region with the similar temperatures \textrm{\citep[see also][]{DeSimone2022}}. This is reasonable given that the sublimation temperature of \methanol and \ammonia are expected to be similar \citep[100--140\,K depending on the gas densities and major composition of ice; e.g.,][]{Furuya14, Hama2013, Minissale2022}. The lower temperature of 4A1 may reflect either higher column density of the dust or lower stellar and accretion luminosities.

The difference in the \dammonia/\ammonia ratios between 4A1 and 4A2 may in fact reflect the difference in the temperature. As discussed in the previous subsections, ices on the grain surface are expected to consist of two layers; the bulk mantle with lower deuteration formed in the molecular cloud stage and the surface component with higher deuteration synthesized in the dense core stage \citep[e.g., ][]{Taquet13, Furuya16, Furuya18}. Thus, the higher \dammonia/\ammonia with a lower temperature in 4A1 may indicate that only the \ammonia ices in the surface layer have sublimated. 

\section{Summary} \label{sec:summary}
We have observed \ammonia and \dammonia high excitation transitions at $\sim$1\arcsec resolution toward the protobinary NGC1333 IRAS4A with the VLA to measure the \dammonia/\ammonia ratio in the warm gas around the protostars, where ammonia ices have sublimated. Our main findings are summarized as follows:

\begin{enumerate}
    \item We have detected the \ammonia (1,1) to (5,5) inversion transitions with S/N of $\gtrsim$10 toward both of the sources, 4A1 and 4A2. We have also detected \dammonia 3$_{1,3}$--3$_{0,3}$ and 4$_{1,4}$--4$_{0,4}$ rotation transitions with S/N of $\sim$5 toward 4A1, while they have been marginally detected with S/N of $\sim$3--\textrm{5} toward 4A2.
    
    \item We have estimated the excitation temperature and column density of \ammonia and \dammonia, and \dammonia/\ammonia column density ratios from the spectral line fitting. We found high \ammonia and \dammonia column densities of $\sim$10$^{17}$--10$^{18}$\,cm$^{-2}$ with high excitation temperatures ($\gtrsim$100\,K) for both 4A1 and 4A2, indicating that the \ammonia and \dammonia lines originate from the inner warm regions. The \dammonia/\ammonia ratios are remarkably high, $\sim$0.5--1. \textrm{The ratio is $\gtrsim$0.3, even if we take into account the slightly higher \ammonia column density derived by \citet{DeSimone2022}.}
    
    \item The high \dammonia/\ammonia ratios indicate the efficient formation of ammonia ices in the prestellar core stage, which in turn suggests that the primary nitrogen reservoir in molecular clouds is atomic nitrogen rather than nitrogen-bearing molecules such as \Ntwo and \ammonia.

    
    
    
    \item In IRAS4A1, the line widths of \dammonia are similar to that of \ammonia (3, 3) transition, but are narrower than those of \ammonia higher transitions. It may indicate a ring-like distribution of \dammonia. The ring-like distribution of \dammonia and very high \dammonia/\ammonia ratio could be due to sublimation of multi-layered ice mantle. The highly deuterated \ammonia ice is formed in the outermost layer of the ice mantle in the late stage of prestellar core and is the first to be sublimated when the grains fall onto the central warm regions.
    
    \item The obtained D/H ratios of ammonia ices have been compared with those of other major volatiles in the IRAS4A core and the comet 67P. The higher D/H ratios of ammonia than those of water and methanol in IRAS4A indicate that ammonia ices are formed in the later stage of the prestellar core phase, where the deuteration is particularly efficient. The higher \dammonia/\ammonia ratio in IRAS4A than that in the comet 67P may suggest that chemical reprocessing during the accretion onto the protostellar disk and subsequent protosplanetary disk phase. Alternative explanation is sublimation of highly deuterated layer of ice mantle in IRAS4A.

\end{enumerate}

\begin{acknowledgments}
\textrm{We thank anonymous referee for valuable comments. We also thank Gianni Cataldi, Satoshi Yamamoto, Nami Sakai, and Yoko Oya for fruitful discussions and comments on the manuscript. We also thank Olli Sipil{\"a} and Shota Notsu for providing the spectroscopic data of \ammonia and a physical model of IRAS 4A which are used for discussions, respectively.} Y.Y. acknowledges support from IGPEES, WINGS Program, the University of Tokyo. K.F. acknowledges support from JSPS KAKENHI grant Nos. 20H05847, 21H01145, and 21K13967. Y.A. acknowledges support by NAOJ ALMA Scientific Research Grant code 2019-13B, Grant-in-Aid for Scientific Research (S) 18H05222, and Grant-in-Aid for Transformative Research Areas (A) 20H05844 and 20H05847. J.K.J. acknowledges support from the Independent Research Fund Denmark (grant No. DFF0135-00123B). The National Radio Astronomy Observatory is a facility of the National Science Foundation operated under cooperative agreement by Associated Universities, Inc.
\end{acknowledgments}

%

\vspace{5mm}
\facilities{VLA}


\software{astropy \citep{2013A&A...558A..33A,2018AJ....156..123A}, bettermoments \citep{Teague18_bettermoments}, astroquery \citep{astroquery}, LIME \citep{LIME}, emcee \citep{emcee}}



\appendix

\section{Hyperfine models for \texorpdfstring{\ammonia}{} and \texorpdfstring{\dammonia}{}}\label{appendix:HFS_model}
Here we describe the details of hyperfine models for observed \ammonia and \dammonia transitions.

\subsection{\texorpdfstring{\ammonia}{} model}
We follow the methodology described in \citet{Rosolowsky08}. Observed spectra in terms of brightness temperature $T_\mathrm{B}(v)$ are modeled as
\begin{equation}\label{eq:RT_model}
    T_\mathrm{B}(v) = f(J_\nu(T_\mathrm{ex}) - J_\nu(T_\mathrm{bg}))(1-e^{-\tau(v)}),
\end{equation}
where $f$ is the beam filling factor as in Equation \ref{eq:bff}, $T_\mathrm{ex}$ is the excitation temperature, $T_\mathrm{bg}$ is the background temperature (i.e., CMB temperature of 2.73\,K here), $\tau(v)$ is the optical depth profile.
For each transition of \ammonia, $\tau(v)$ is given by
\begin{equation}
    \tau(v) = \tau_0\,\sum_i r_i \exp\left(-\frac{(v-\delta v_i-v_0)^2}{2\sigma_{v,i}^2}\right),
\end{equation}
where $\tau_0$ is the total optical depth of the transition, $r_i$, $\delta v_i$, and $\sigma_{v, i}$ are the relative strength, velocity offset, and line width of $i$th hyperfine component, and $v_0$ is the systemic velocity of the central protostar. The values of $r_i$, $\delta v_i$, and $\sigma_{v, i}$ are calculated as
\begin{equation}
    r_i = \frac{g_{\mathrm{u},i}A_{\mathrm{ul},i}}{\sum_j g_{\mathrm{u},j}A_{\mathrm{u},j}}, \quad
    \delta v_i = \left(1 - \frac{\nu_i}{\nu_0}\right)c, \quad
    \sigma_{v,i} = \frac{\sigma_v}{c}\nu_i,
\end{equation}
where $\nu_0$ is the rest frequency of inversion transition listed in Table \ref{tab:observed_transitions}, $g_{\mathrm{u}, i}$, $A_{\mathrm{ul}, i}$, and $\nu_i$ are the statistical weight, Einstein A coefficient for spontaneous emission, and the rest frequency of $i$th hyperfine component, respectively \textrm{\citep[Table \ref{tab:HFS_data},][]{Kukolich67, Kukolich70}}. These data are retrieved from the Cologne Database for Molecular Spectroscopy \citep[CDMS,][]{CDMS} available at the Splatalogue database queried by \texttt{astroquery} \citep{astroquery}. We assumed that the $\sigma_v$ and $v_0$ are common among the hyperfine components in the same inversion or rotational transition, given that they originate from the same temperature region.


The total optical depths of each transition are related back to the column density as in \citet{Friesen09} \citep[see also][]{Rosolowsky08};
\begin{equation}
    N(J,K) = \sqrt{2\pi}\sigma_v\frac{8\pi\nu_0^3}{c^3A_\mathrm{ul}}\frac{1 + \exp(-h\nu_0/k_\mathrm{B}T_\mathrm{ex})}{1 - \exp(-h\nu_0/k_\mathrm{B}T_\mathrm{ex})}\tau_0(J,K)
\end{equation}
where $N(J, K)$ is the column density of \ammonia at the $(J, K)$ metastable state. The column density $N(J, K)$ can be related with the total (ortho or para) \ammonia column density $N$(\ammonia) through the partition function $Q$ as
\begin{equation}
    N(\mathrm{NH_3}) = \frac{Q(T_\mathrm{rot})}{g_\mathrm{u}}N(J,K)\exp\left(\frac{E_\mathrm{u}}{T_\mathrm{rot}}\right)
\end{equation}
where $g_\mathrm{u}$ and $E_\mathrm{u}$ are the statistical weight and energy of the upper state listed in Table \ref{tab:observed_transitions}. The partition function $Q$ is calculated following \citet{Wilson09_ToRA} as
\begin{equation}
    Q(T) = \sum_{J=0}^{\infty}\sum_{K=0}^{K=J}(2J+1)S(J,K)\exp\left[-\frac{h(BJ(J+1) + (C-B)K^2)}{k_\mathrm{B}T}\right], 
\end{equation}
where $B=298117$\,MHz and $C=186726$\,MHz are the rotational constants of \ammonia, $S(J, K)$ is the constant factor for the nuclear spin statistics. For ortho- and para-\ammonia transitions, $S(J,K) = 4, 2$, respectively. 

In the LTE approximation, the excitation temperature ($T_\mathrm{ex}$) that govern the inversion transition and the rotational temperature ($T_\mathrm{rot}$) that govern the population between the metastable levels are assumed to be the same, and these temperatures can be directly used as a proxy of the kinetic temperature ($T_\mathrm{kin}$). In our fit, all of these temperatures are assumed to be the same and represented by $T_\mathrm{ex}$.

\subsection{\texorpdfstring{\dammonia}{} model}
Our \dammonia model follows the same formulation as \ammonia. The spectroscopic data used for \dammonia are listed in Table \ref{tab:HFS_data} \textrm{\citep[][]{DeLucia75, Cohen82, Fusina88}}. Since we observed only single transition for each ortho- and para-\dammonia, we directly related the optical depth to the column density assuming that $T_\mathrm{ex}$ is the same as that of \ammonia; 
\begin{equation}
    N_\mathrm{u} = \sqrt{2\pi}\sigma_v\frac{8\pi\nu_0^3}{c^3A_\mathrm{ul}}\left[\exp\left(\frac{h\nu_0}{k_\mathrm{B}T_\mathrm{ex}}\right) - 1\right]\tau_0
\end{equation}
where $N_\mathrm{u}$ is the \dammonia column density at the upper state of the transition. The total \dammonia column density is calculated as
\begin{equation}
    N(\mathrm{NH_2D}) = \frac{Q(T_\mathrm{ex})}{g_\mathrm{u}}N_u\exp\left(\frac{E_\mathrm{u}}{T_\mathrm{ex}}\right),
\end{equation}
where $Q$ is the full (i.e., ortho+para) partition function of \dammonia retrieved from the CDMS \citep{CDMS}.

\begin{deluxetable}{ccCCCCC}
\label{tab:HFS_data}
\tablecaption{Spectroscopic Data of \ammonia and \dammonia Used in Hyperfine Fits}
\tablehead{\colhead{Transition} & \colhead{$F$} & \colhead{$\nu_0$} & \colhead{$g_\mathrm{u}$} & \colhead{$\log_{10}A_\mathrm{ul}$} & \colhead{$r^\dagger$} & \colhead{$\delta v^\ddagger$} \\
\colhead{} & \colhead{} & \colhead{[GHz]} & \colhead{} & \colhead{[s$^{-1}$]} & \colhead{} & \colhead{[km\,s$^{-1}$]}}
\startdata
NH$_3$ $(3,3)$ & 2-3 & 23.8678062 & 20.0 & -7.5449 & 0.0264 & 29.1753 \\
 & 4-3 & 23.8684499 & 36.0 & -7.794 & 0.0268 & 21.0909 \\
 & 3-3 & 23.8701279 & 28.0 & -6.6658 & 0.28 & 0.0163 \\
 & 4-4 & 23.8701296 & 36.0 & -6.6174 & 0.4024 & -0.005 \\
 & 2-2 & 23.8701303 & 20.0 & -6.642 & 0.2112 & -0.0138 \\
 & 3-4 & 23.8718076 & 28.0 & -7.6846 & 0.0268 & -21.0796 \\
 & 3-2 & 23.872452 & 28.0 & -7.6908 & 0.0264 & -29.1728 \\
\hline
NH$_3$ $(4,4)$ & 3-4 & 24.1369314 & 14.0 & -7.7529 & 0.0162 & 30.8605 \\
 & 5-4 & 24.1374943 & 22.0 & -7.9457 & 0.0163 & 23.8697 \\
 & 4-4 & 24.139415 & 18.0 & -6.5927 & 0.3007 & 0.0161 \\
 & 5-5 & 24.139417 & 22.0 & -6.5648 & 0.3919 & -0.0087 \\
 & 3-3 & 24.1394175 & 14.0 & -6.5771 & 0.2424 & -0.0149 \\
 & 4-5 & 24.1413376 & 18.0 & -7.8583 & 0.0163 & -23.861 \\
 & 4-3 & 24.1419011 & 18.0 & -7.8617 & 0.0162 & -30.8593 \\
\hline
NH$_3$ $(5,5)$ & 4-5 & 24.5303911 & 18.0 & -7.9088 & 0.0109 & 31.7426 \\
 & 6-5 & 24.5308986 & 26.0 & -8.0659 & 0.011 & 25.541 \\
 & 5-5 & 24.532987 & 22.0 & -6.5396 & 0.3113 & 0.0208 \\
 & 6-6 & 24.5329892 & 26.0 & -6.5211 & 0.384 & -0.0061 \\
 & 4-4 & 24.5329897 & 18.0 & -6.529 & 0.261 & -0.0122 \\
 & 5-6 & 24.5350777 & 22.0 & -7.9932 & 0.011 & -25.5275 \\
 & 5-4 & 24.5355856 & 22.0 & -7.9957 & 0.0109 & -31.734 \\
\hline
NH$_2$D 3$_{1,3}$--3$_{0,3}$ & 2-3 & 18.805907 & 15.0 & -8.3826 & 0.0265 & 31.5766 \\
 & 4-3 & 18.80649 & 27.0 & -8.6324 & 0.0268 & 22.2837 \\
 & 2-2 & 18.807831 & 15.0 & -7.4793 & 0.2116 & 0.9086 \\
 & 2-2 & 18.807831 & 27.0 & -7.4562 & 0.4018 & 0.9086 \\
 & 3-3 & 18.8080466 & 21.0 & -7.5037 & 0.2801 & -2.528 \\
 & 3-4 & 18.809385 & 21.0 & -8.523 & 0.0268 & -23.8618 \\
 & 3-2 & 18.8099083 & 21.0 & -8.5284 & 0.0265 & -32.203 \\
\hline
NH$_2$D 4$_{1,4}$--4$_{0,4}$ & 3-4 & 25.0216041 & 7.0 & -8.2179 & 0.0162 & 26.2117 \\
 & 5-4 & 25.0221167 & 11.0 & -8.4117 & 0.0163 & 20.0706 \\
 & 3-3 & 25.0237541 & 7.0 & -7.0417 & 0.2431 & 0.4541 \\
 & 5-5 & 25.0237746 & 11.0 & -7.0314 & 0.3911 & 0.2085 \\
 & 4-4 & 25.0238552 & 9.0 & -7.0582 & 0.3008 & -0.7572 \\
 & 4-5 & 25.0255132 & 9.0 & -8.3244 & 0.0163 & -20.6205 \\
 & 4-3 & 25.0260052 & 9.0 & -8.3268 & 0.0162 & -26.5148 \\
\enddata
\tablenotetext{}{$\dagger$ Relative strengths of hyperfine transitions. Normalized as the sum of $r$ being one.}
\tablenotetext{}{$\ddagger$ Velocity offsets with respect to the inversion or rotation transitions listed in Table \ref{tab:observed_transitions}.}
\end{deluxetable}

\section{Radiative transfer modeling}\label{appendix:RT_model}

\begin{figure}
\epsscale{0.6}
\plotone{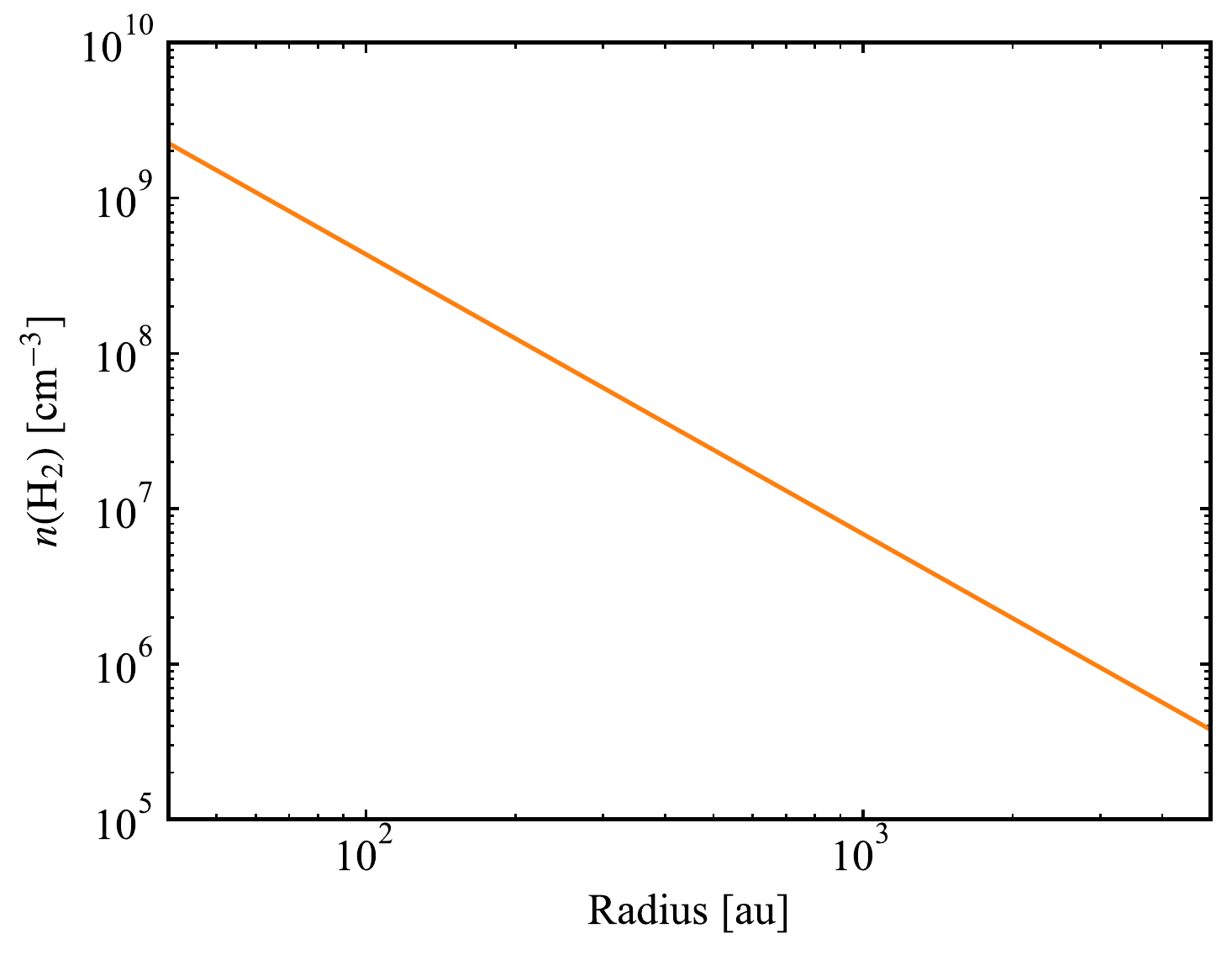}
\caption{Gas density profile of the IRAS4A envelope presented in \citet{Persson16}.}
\label{fig:IRAS4A_model_nH2}
\end{figure}

\begin{figure}
\epsscale{0.6}
\plotone{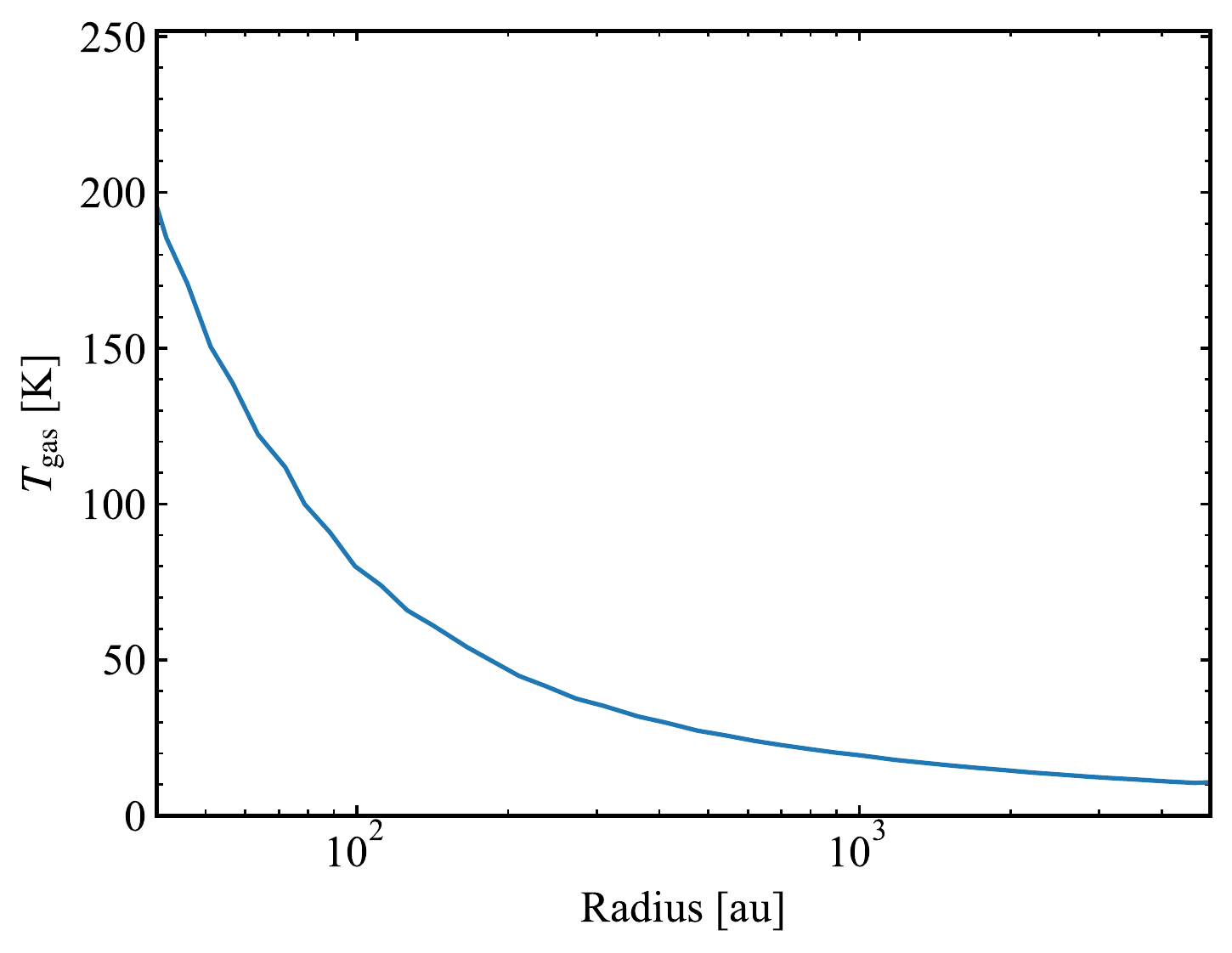}
\caption{Dust temperature profile of the IRAS4A envelope presented in \citet{Persson16}.}
\label{fig:IRAS4A_model_T}
\end{figure}

We simulated the spectra of \dammonia transitions using the non-LTE radiative transfer code LIME \citep{LIME} to see if only the cold envelope component of \dammonia can reproduce the observed \dammonia emission. Here we will describe the simulation settings. As the density and temperature structures of the IRAS4A protostellar cores, we used the spherically symmetric envelope model presented in \citet{Persson16}. The gas density and dust temperature profiles are shown in Figure \ref{fig:IRAS4A_model_nH2} and Figure \ref{fig:IRAS4A_model_T}. The gas density follows a single power-law profile, $n = n_\mathrm{in}\left(r/r_\mathrm{in}\right)^{-p_\mathrm{env}}$, where the parameters of $n_\mathrm{in} = 3.1\times10^9\,$cm$^{-3}$, $r_\mathrm{in}=33.5$\,au, and $p_\mathrm{env}=1.8$ are constrained through the simultaneous fits to sub-mm dust continuum radial profiles and spectral energy distributions (SEDs) \citep[see also][]{Jorgensen02, Kristensen12}. For this profile of the gas density, the dust temperature profile is self-consistently calculated \citep{Jorgensen02, Kristensen12, Persson16} assuming the observed luminosity of $9.1\,L_\odot$. We adopted this dust temperature profile as the gas temperature profile, i.e., assumed that the gas and dust are well-coupled and their temperatures are the same. The modeled dust temperature inside the observed emitting region size ($\sim$75\,au) is $\gtrsim100$\,K, consistent with the observed excitation temperatures of $\sim100$\,K and $\sim160$\,K (see Section \ref{sec:results}). The radial profiles are truncated at $r_\mathrm{in}$ and $r_\mathrm{out} = 33500$\,au.

For the kinetics of the gas, we simply employed the spherically symmetric free-falling envelope with a radial velocity of $v_\mathrm{ff} = \sqrt{2GM_\star/r}$. We assumed the protostellar mass of $M_\star=1M_\odot$ for both 4A1 and 4A2, given that there are almost no measurement of the central stellar mass. For 4A2, \citet{Choi10} estimated it to be 0.08\,$M_\odot$ from the velocity gradient analysis of \ammonia emission. Thus, we took another run with $M_\star = 0.08\,M_\odot$ to evaluate the effect of the central stellar mass to observed emission. 

We calculated the \dammonia abundance in the cold envelope based on the single-dish observations and the model H$_2$ column density. \citet{Hatchell03} observed the \dammonia rotation-inversion transition with the IRAM 30\,m telescope (with a beam size of $\sim$25\arcsec or $\sim$7000\,au), and derived an \dammonia column density of $3.9\times10^{14}$\,cm$^{-2}$. This value is divided by the model H$_2$ column density of $1.9\times10^{24}$\,cm$^{-2}$ \citep[see Table C.1 in][]{Kristensen12}, resulting in an \dammonia abundance of 2.1
$\times$10$^{-10}$. A constant abundance with radius is employed. The ortho-to-para ratio of \dammonia is fixed to three, as done for the observational data analysis (Section \ref{sec:analysis}).

The collisional excitation rates of ortho-\dammonia and para-\dammonia are based on \citet{Daniel14} and taken from the Leiden Atomic and Molecular Database \citep[LAMDA,][]{LAMDA}. Given that the collisional excitation rates for ortho-H$_2$ and para-H$_2$ are available, we employ the thermal ortho-to-para ratio of H$_2$ \citep{Flower84, Flower85};
\begin{equation}
    \frac{n(\mathrm{ortho\mathchar`-H_2})}{n(\mathrm{para\mathchar`-H_2})} = 9 \exp\left(-\frac{170.5\,\mathrm{K}}{T}\right),
\end{equation}
where $T$ is the gas temperature. This ratio is low at low temperature, e.g., $\sim$0.03 at 30\,K. 

We use the dust opacity with thin ice mantle presented in the fifth columns of Table~1 in \citet{Ossenkopf94} and a standard gas-to-dust mass ratio of 100. The dust opacity is the same as the one used in the modeling by \citet{Persson16}.

We run the LIME code with its non-LTE mode to produce the model image cubes of \dammonia $3_{1,3}-3_{0,3}$ and $4_{1,4}-4_{0,4}$ transitions. To simulate the observations, first we convolved the output image cubes by the 2D Gaussian with the FWHM of the beam major and minor axis. 
Then, we subtracted the continuum emission by the CASA task \textit{imcontsub} and extracted the spectra toward the continuum peak by the same manner as the case of the observations. 

\begin{figure*}
\plotone{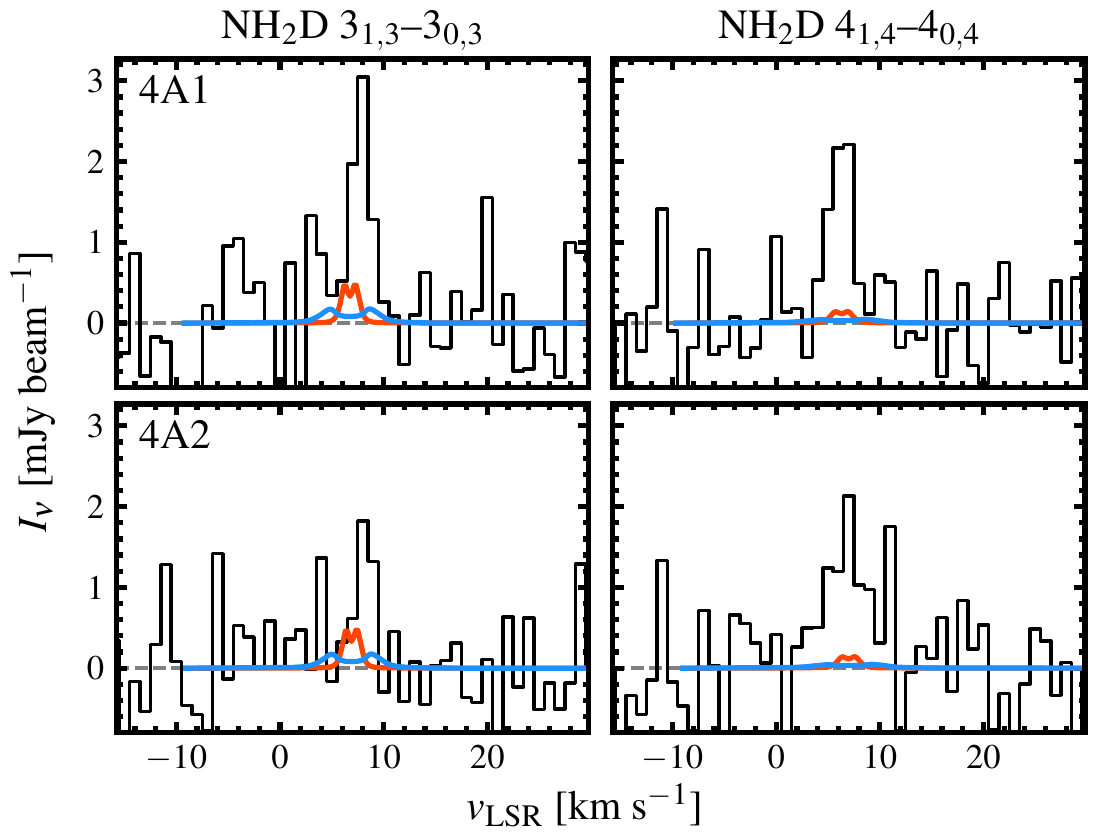}
\caption{Comparisom of model spectra of \dammonia transitions predicted by the non-LTE simulation with central stellar masses of $1.0\,M_\odot$ (blue) and $0.08\,M_\odot$ (orange) and observed spectra toward 4A1 (top) and 4A2 (bottom). The zero intensity level is indicated by the grey dashed line.}
\label{fig:NH2D_LIME_spectra}
\end{figure*}

The simulated spectra are compared to the observed spectra in Figure \ref{fig:NH2D_LIME_spectra}. The model with 0.08\,$M_\odot$ stellar mass show the brighter peak intensity due to the narrower line width caused by lower free-falling velocities. For both models, the peak intensity of the model spectra are weaker compared to the observed spectra by a factor of $\gtrsim$10, except for \dammonia 3$_{1,3}$--3$_{0,3}$ with a central stellar mass of 0.08\,$M_\odot$. The cold component may partially contribute to \dammonia 3$_{1,3}$--3$_{0,3}$ emission ($\sim$20\% in velocity-integrated intensity) if the central stellar mass is small. We note that, however, the hyperfine splitting is not considered in this modeling. If it is considered, the peak intensity can be weaker because the total emission will be distributed over the hyperfine satellites. This is particularly effective for \dammonia 3$_{1,3}$--3$_{0,3}$ where it has a prominent splitting around the main component (see Table \ref{tab:HFS_data} for the hyperfine spectroscopic data). Thus, this comparison between model and observed spectra suggests that the observed emission cannot be reproduced by the cold component alone. Therefore, we suggest that the observed \dammonia emission indeed traces the \dammonia sublimated from ice in the hot corino rather than the cold envelope component, and the degree of \ammonia ice deuteration is indeed high in the protostellar cores of IRAS 4A1 and 4A2. We emphasize that the purpose of the radiative transfer modeling presented here is to just evaluate the contributions of the cold component of \dammonia in the envelope to the observed emission. More detailed modeling is needed to reproduce the observed emission. Although the detailed physical structure on smaller scales can differ from the simple free-falling envelope model presented here \citep[e.g., protostellar disk as considered by][]{Choi10}, it is beyond the scope of the present work.

\bibliography{sample631}{}

\begin{thebibliography}{}
\expandafter\ifx\csname natexlab\endcsname\relax\def\natexlab#1{#1}\fi
\providecommand{\url}[1]{\href{#1}{#1}}
\providecommand{\dodoi}[1]{doi:~\href{http://doi.org/#1}{\nolinkurl{#1}}}
\providecommand{\doeprint}[1]{\href{http://ascl.net/#1}{\nolinkurl{http://ascl.net/#1}}}
\providecommand{\doarXiv}[1]{\href{https://arxiv.org/abs/#1}{\nolinkurl{https://arxiv.org/abs/#1}}}

\bibitem[{{Aikawa} {et~al.}(2005){Aikawa}, {Herbst}, {Roberts}, \&
  {Caselli}}]{Aikawa2005}
{Aikawa}, Y., {Herbst}, E., {Roberts}, H., \& {Caselli}, P. 2005, \apj, 620,
  330, \dodoi{10.1086/427017}

\bibitem[{{Aikawa} {et~al.}(2012){Aikawa}, {Wakelam}, {Hersant}, {Garrod}, \&
  {Herbst}}]{Aikawa12}
{Aikawa}, Y., {Wakelam}, V., {Hersant}, F., {Garrod}, R.~T., \& {Herbst}, E.
  2012, \apj, 760, 40, \dodoi{10.1088/0004-637X/760/1/40}

\bibitem[{{Altwegg} {et~al.}(2019){Altwegg}, {Balsiger}, \&
  {Fuselier}}]{Altwegg19}
{Altwegg}, K., {Balsiger}, H., \& {Fuselier}, S.~A. 2019, \araa, 57, 113,
  \dodoi{10.1146/annurev-astro-091918-104409}

\bibitem[{{Altwegg} {et~al.}(2017){Altwegg}, {Balsiger}, {Berthelier},
  {Bieler}, {Calmonte}, {De Keyser}, {Fiethe}, {Fuselier}, {Gasc}, {Gombosi},
  {Owen}, {Le Roy}, {Rubin}, {S{\'e}mon}, \& {Tzou}}]{Altwegg17}
{Altwegg}, K., {Balsiger}, H., {Berthelier}, J.~J., {et~al.} 2017,
  Philosophical Transactions of the Royal Society of London Series A, 375,
  20160253, \dodoi{10.1098/rsta.2016.0253}

\bibitem[{{Altwegg} {et~al.}(2020){Altwegg}, {Balsiger}, {H{\"a}nni}, {Rubin},
  {Schuhmann}, {Schroeder}, {S{\'e}mon}, {Wampfler}, {Berthelier}, {Briois},
  {Combi}, {Gombosi}, {Cottin}, {De Keyser}, {Dhooghe}, {Fiethe}, \&
  {Fuselier}}]{Altwegg20}
{Altwegg}, K., {Balsiger}, H., {H{\"a}nni}, N., {et~al.} 2020, Nature
  Astronomy, 4, 533, \dodoi{10.1038/s41550-019-0991-9}

\bibitem[{{Astropy Collaboration} {et~al.}(2013){Astropy Collaboration},
  {Robitaille}, {Tollerud}, {Greenfield}, {Droettboom}, {Bray}, {Aldcroft},
  {Davis}, {Ginsburg}, {Price-Whelan}, {Kerzendorf}, {Conley}, {Crighton},
  {Barbary}, {Muna}, {Ferguson}, {Grollier}, {Parikh}, {Nair}, {Unther},
  {Deil}, {Woillez}, {Conseil}, {Kramer}, {Turner}, {Singer}, {Fox}, {Weaver},
  {Zabalza}, {Edwards}, {Azalee Bostroem}, {Burke}, {Casey}, {Crawford},
  {Dencheva}, {Ely}, {Jenness}, {Labrie}, {Lim}, {Pierfederici}, {Pontzen},
  {Ptak}, {Refsdal}, {Servillat}, \& {Streicher}}]{2013A&A...558A..33A}
{Astropy Collaboration}, {Robitaille}, T.~P., {Tollerud}, E.~J., {et~al.} 2013,
  \aap, 558, A33, \dodoi{10.1051/0004-6361/201322068}

\bibitem[{{Astropy Collaboration} {et~al.}(2018){Astropy Collaboration},
  {Price-Whelan}, {Sip{\H{o}}cz}, {G{\"u}nther}, {Lim}, {Crawford}, {Conseil},
  {Shupe}, {Craig}, {Dencheva}, {Ginsburg}, {VanderPlas}, {Bradley},
  {P{\'e}rez-Su{\'a}rez}, {de Val-Borro}, {Aldcroft}, {Cruz}, {Robitaille},
  {Tollerud}, {Ardelean}, {Babej}, {Bach}, {Bachetti}, {Bakanov}, {Bamford},
  {Barentsen}, {Barmby}, {Baumbach}, {Berry}, {Biscani}, {Boquien}, {Bostroem},
  {Bouma}, {Brammer}, {Bray}, {Breytenbach}, {Buddelmeijer}, {Burke},
  {Calderone}, {Cano Rodr{\'\i}guez}, {Cara}, {Cardoso}, {Cheedella}, {Copin},
  {Corrales}, {Crichton}, {D'Avella}, {Deil}, {Depagne}, {Dietrich}, {Donath},
  {Droettboom}, {Earl}, {Erben}, {Fabbro}, {Ferreira}, {Finethy}, {Fox},
  {Garrison}, {Gibbons}, {Goldstein}, {Gommers}, {Greco}, {Greenfield},
  {Groener}, {Grollier}, {Hagen}, {Hirst}, {Homeier}, {Horton}, {Hosseinzadeh},
  {Hu}, {Hunkeler}, {Ivezi{\'c}}, {Jain}, {Jenness}, {Kanarek}, {Kendrew},
  {Kern}, {Kerzendorf}, {Khvalko}, {King}, {Kirkby}, {Kulkarni}, {Kumar},
  {Lee}, {Lenz}, {Littlefair}, {Ma}, {Macleod}, {Mastropietro}, {McCully},
  {Montagnac}, {Morris}, {Mueller}, {Mumford}, {Muna}, {Murphy}, {Nelson},
  {Nguyen}, {Ninan}, {N{\"o}the}, {Ogaz}, {Oh}, {Parejko}, {Parley}, {Pascual},
  {Patil}, {Patil}, {Plunkett}, {Prochaska}, {Rastogi}, {Reddy Janga},
  {Sabater}, {Sakurikar}, {Seifert}, {Sherbert}, {Sherwood-Taylor}, {Shih},
  {Sick}, {Silbiger}, {Singanamalla}, {Singer}, {Sladen}, {Sooley},
  {Sornarajah}, {Streicher}, {Teuben}, {Thomas}, {Tremblay}, {Turner},
  {Terr{\'o}n}, {van Kerkwijk}, {de la Vega}, {Watkins}, {Weaver}, {Whitmore},
  {Woillez}, {Zabalza}, \& {Astropy Contributors}}]{2018AJ....156..123A}
{Astropy Collaboration}, {Price-Whelan}, A.~M., {Sip{\H{o}}cz}, B.~M., {et~al.}
  2018, \aj, 156, 123, \dodoi{10.3847/1538-3881/aabc4f}

\bibitem[{{Bianchi} {et~al.}(2019){Bianchi}, {Codella}, {Ceccarelli}, {Vazart},
  {Bachiller}, {Balucani}, {Bouvier}, {De Simone}, {Enrique-Romero}, {Kahane},
  {Lefloch}, {L{\'o}pez-Sepulcre}, {Ospina-Zamudio}, {Podio}, \&
  {Taquet}}]{Bianchi19}
{Bianchi}, E., {Codella}, C., {Ceccarelli}, C., {et~al.} 2019, \mnras, 483,
  1850, \dodoi{10.1093/mnras/sty2915}

\bibitem[{{Boogert} {et~al.}(2015){Boogert}, {Gerakines}, \&
  {Whittet}}]{Boogert15}
{Boogert}, A.~C.~A., {Gerakines}, P.~A., \& {Whittet}, D. C.~B. 2015, \araa,
  53, 541, \dodoi{10.1146/annurev-astro-082214-122348}

\bibitem[{{Bottinelli} {et~al.}(2010){Bottinelli}, {Boogert}, {Bouwman},
  {Beckwith}, {van Dishoeck}, {{\"O}berg}, {Pontoppidan}, {Linnartz}, {Blake},
  {Evans}, \& {Lahuis}}]{Bottinelli10}
{Bottinelli}, S., {Boogert}, A.~C.~A., {Bouwman}, J., {et~al.} 2010, \apj, 718,
  1100, \dodoi{10.1088/0004-637X/718/2/1100}

\bibitem[{{Brinch} \& {Hogerheijde}(2010)}]{LIME}
{Brinch}, C., \& {Hogerheijde}, M.~R. 2010, \aap, 523, A25,
  \dodoi{10.1051/0004-6361/201015333}

\bibitem[{{Caselli} {et~al.}(2022){Caselli}, {Pineda}, {Sipil{\"a}}, {Zhao},
  {Redaelli}, {Spezzano}, {Maureira}, {Alves}, {Bizzocchi}, {Bourke},
  {Chac{\'o}n-Tanarro}, {Friesen}, {Galli}, {Harju}, {Jim{\'e}nez-Serra},
  {Keto}, {Li}, {Padovani}, {Schmiedeke}, {Tafalla}, \& {Vastel}}]{Caselli2022}
{Caselli}, P., {Pineda}, J.~E., {Sipil{\"a}}, O., {et~al.} 2022, \apj, 929, 13,
  \dodoi{10.3847/1538-4357/ac5913}

\bibitem[{{Choi} {et~al.}(2011){Choi}, {Kang}, {Tatematsu}, {Lee}, \&
  {Park}}]{Choi11}
{Choi}, M., {Kang}, M., {Tatematsu}, K., {Lee}, J.-E., \& {Park}, G. 2011,
  \pasj, 63, 1281, \dodoi{10.1093/pasj/63.6.1281}

\bibitem[{{Choi} {et~al.}(2010){Choi}, {Tatematsu}, \& {Kang}}]{Choi10}
{Choi}, M., {Tatematsu}, K., \& {Kang}, M. 2010, \apjl, 723, L34,
  \dodoi{10.1088/2041-8205/723/1/L34}

\bibitem[{{Choi} {et~al.}(2007){Choi}, {Tatematsu}, {Park}, \& {Kang}}]{Choi07}
{Choi}, M., {Tatematsu}, K., {Park}, G., \& {Kang}, M. 2007, \apjl, 667, L183,
  \dodoi{10.1086/522116}

\bibitem[{{Cohen} \& {Pickett}(1982)}]{Cohen82}
{Cohen}, E.~A., \& {Pickett}, H.~M. 1982, Journal of Molecular Spectroscopy,
  93, 83, \dodoi{10.1016/0022-2852(82)90276-4}

\bibitem[{{Daniel} {et~al.}(2014){Daniel}, {Faure}, {Wiesenfeld}, {Roueff},
  {Lis}, \& {Hily-Blant}}]{Daniel14}
{Daniel}, F., {Faure}, A., {Wiesenfeld}, L., {et~al.} 2014, \mnras, 444, 2544,
  \dodoi{10.1093/mnras/stu1670}

\bibitem[{{De Lucia} \& {Helminger}(1975)}]{DeLucia75}
{De Lucia}, F.~C., \& {Helminger}, P. 1975, Journal of Molecular Spectroscopy,
  54, 200, \dodoi{10.1016/0022-2852(75)90074-0}

\bibitem[{{De Simone} {et~al.}(2020){De Simone}, {Ceccarelli}, {Codella},
  {Svoboda}, {Chandler}, {Bouvier}, {Yamamoto}, {Sakai}, {Caselli}, {Favre},
  {Loinard}, {Lefloch}, {Liu}, {L{\'o}pez-Sepulcre}, {Pineda}, {Taquet}, \&
  {Testi}}]{DeSimone20}
{De Simone}, M., {Ceccarelli}, C., {Codella}, C., {et~al.} 2020, \apjl, 896,
  L3, \dodoi{10.3847/2041-8213/ab8d41}

\bibitem[{{De Simone} {et~al.}(2022){De Simone}, {Ceccarelli}, {Codella},
  {Svoboda}, {Chandler}, {Bouvier}, {Yamamoto}, {Sakai}, {Yang}, {Caselli},
  {Lefloch}, {Liu}, {L{\'o}pez-Sepulcre}, {Loinard}, {Pineda}, \&
  {Testi}}]{DeSimone2022}
---. 2022, arXiv e-prints, arXiv:2208.00247.
\newblock \doarXiv{2208.00247}

\bibitem[{{Di Francesco} {et~al.}(2001){Di Francesco}, {Myers}, {Wilner},
  {Ohashi}, \& {Mardones}}]{DiFrancesco01}
{Di Francesco}, J., {Myers}, P.~C., {Wilner}, D.~J., {Ohashi}, N., \&
  {Mardones}, D. 2001, \apj, 562, 770, \dodoi{10.1086/323854}

\bibitem[{{Drozdovskaya} {et~al.}(2019){Drozdovskaya}, {van Dishoeck}, {Rubin},
  {J{\o}rgensen}, \& {Altwegg}}]{Drozdovskaya19}
{Drozdovskaya}, M.~N., {van Dishoeck}, E.~F., {Rubin}, M., {J{\o}rgensen},
  J.~K., \& {Altwegg}, K. 2019, \mnras, 490, 50, \dodoi{10.1093/mnras/stz2430}

\bibitem[{{Drozdovskaya} {et~al.}(2021){Drozdovskaya}, {Schroeder I}, {Rubin},
  {Altwegg}, {van Dishoeck}, {Kulterer}, {De Keyser}, {Fuselier}, \&
  {Combi}}]{Drozdovskaya21}
{Drozdovskaya}, M.~N., {Schroeder I}, I. R.~H.~G., {Rubin}, M., {et~al.} 2021,
  \mnras, 500, 4901, \dodoi{10.1093/mnras/staa3387}

\bibitem[{{Fedoseev} {et~al.}(2015){Fedoseev}, {Ioppolo}, {Zhao}, {Lamberts},
  \& {Linnartz}}]{Fedoseev2015}
{Fedoseev}, G., {Ioppolo}, S., {Zhao}, D., {Lamberts}, T., \& {Linnartz}, H.
  2015, \mnras, 446, 439, \dodoi{10.1093/mnras/stu2028}

\bibitem[{{Flower} \& {Watt}(1984)}]{Flower84}
{Flower}, D.~R., \& {Watt}, G.~D. 1984, \mnras, 209, 25,
  \dodoi{10.1093/mnras/209.1.25}

\bibitem[{{Flower} \& {Watt}(1985)}]{Flower85}
---. 1985, \mnras, 213, 991, \dodoi{10.1093/mnras/213.4.991}

\bibitem[{{Foreman-Mackey} {et~al.}(2013){Foreman-Mackey}, {Hogg}, {Lang}, \&
  {Goodman}}]{emcee}
{Foreman-Mackey}, D., {Hogg}, D.~W., {Lang}, D., \& {Goodman}, J. 2013, \pasp,
  125, 306, \dodoi{10.1086/670067}

\bibitem[{{Friesen} {et~al.}(2009){Friesen}, {Di Francesco}, {Shirley}, \&
  {Myers}}]{Friesen09}
{Friesen}, R.~K., {Di Francesco}, J., {Shirley}, Y.~L., \& {Myers}, P.~C. 2009,
  \apj, 697, 1457, \dodoi{10.1088/0004-637X/697/2/1457}

\bibitem[{{Furuya} \& {Aikawa}(2014)}]{Furuya14}
{Furuya}, K., \& {Aikawa}, Y. 2014, \apj, 790, 97,
  \dodoi{10.1088/0004-637X/790/2/97}

\bibitem[{{Furuya} {et~al.}(2017){Furuya}, {Drozdovskaya}, {Visser}, {van
  Dishoeck}, {Walsh}, {Harsono}, {Hincelin}, \& {Taquet}}]{Furuya17}
{Furuya}, K., {Drozdovskaya}, M.~N., {Visser}, R., {et~al.} 2017, \aap, 599,
  A40, \dodoi{10.1051/0004-6361/201629269}

\bibitem[{{Furuya} \& {Persson}(2018)}]{Furuya18}
{Furuya}, K., \& {Persson}, M.~V. 2018, \mnras, 476, 4994,
  \dodoi{10.1093/mnras/sty553}

\bibitem[{{Furuya} {et~al.}(2016){Furuya}, {van Dishoeck}, \&
  {Aikawa}}]{Furuya16}
{Furuya}, K., {van Dishoeck}, E.~F., \& {Aikawa}, Y. 2016, \aap, 586, A127,
  \dodoi{10.1051/0004-6361/201527579}

\bibitem[{{Furuya} {et~al.}(2018){Furuya}, {Watanabe}, {Sakai}, {Aikawa}, \&
  {Yamamoto}}]{Furuya18_L1544}
{Furuya}, K., {Watanabe}, Y., {Sakai}, T., {Aikawa}, Y., \& {Yamamoto}, S.
  2018, \aap, 615, L16, \dodoi{10.1051/0004-6361/201833607}

\bibitem[{{Fusina} {et~al.}(1988){Fusina}, {Di Lonardo}, {Johns}, \&
  {Halonen}}]{Fusina88}
{Fusina}, L., {Di Lonardo}, G., {Johns}, J.~W.~C., \& {Halonen}, L. 1988,
  Journal of Molecular Spectroscopy, 127, 240,
  \dodoi{10.1016/0022-2852(88)90023-9}

\bibitem[{{Gibb} {et~al.}(2000){Gibb}, {Whittet}, {Schutte}, {Boogert},
  {Chiar}, {Ehrenfreund}, {Gerakines}, {Keane}, {Tielens}, {van Dishoeck}, \&
  {Kerkhof}}]{Gibb2000}
{Gibb}, E.~L., {Whittet}, D.~C.~B., {Schutte}, W.~A., {et~al.} 2000, \apj, 536,
  347, \dodoi{10.1086/308940}

\bibitem[{{Ginsburg} {et~al.}(2019){Ginsburg}, {Sip{\H{o}}cz}, {Brasseur},
  {Cowperthwaite}, {Craig}, {Deil}, {Guillochon}, {Guzman}, {Liedtke}, {Lian
  Lim}, {Lockhart}, {Mommert}, {Morris}, {Norman}, {Parikh}, {Persson},
  {Robitaille}, {Segovia}, {Singer}, {Tollerud}, {de Val-Borro}, {Valtchanov},
  {Woillez}, {Astroquery Collaboration}, \& {a subset of astropy
  Collaboration}}]{astroquery}
{Ginsburg}, A., {Sip{\H{o}}cz}, B.~M., {Brasseur}, C.~E., {et~al.} 2019, \aj,
  157, 98, \dodoi{10.3847/1538-3881/aafc33}

\bibitem[{{Hama} \& {Watanabe}(2013)}]{Hama2013}
{Hama}, T., \& {Watanabe}, N. 2013, Chemical Reviews, 113, 8783,
  \dodoi{10.1021/cr4000978}

\bibitem[{{Hatchell}(2003)}]{Hatchell03}
{Hatchell}, J. 2003, \aap, 403, L25, \dodoi{10.1051/0004-6361:20030297}

\bibitem[{{Hidaka} {et~al.}(2011){Hidaka}, {Watanabe}, {Kouchi}, \&
  {Watanabe}}]{Hidaka2011}
{Hidaka}, H., {Watanabe}, M., {Kouchi}, A., \& {Watanabe}, N. 2011, Physical
  Chemistry Chemical Physics (Incorporating Faraday Transactions), 13, 15798,
  \dodoi{10.1039/C1CP20645A}

\bibitem[{{Jensen} {et~al.}(2019){Jensen}, {J{\o}rgensen}, {Kristensen},
  {Furuya}, {Coutens}, {van Dishoeck}, {Harsono}, \& {Persson}}]{Jensen19}
{Jensen}, S.~S., {J{\o}rgensen}, J.~K., {Kristensen}, L.~E., {et~al.} 2019,
  \aap, 631, A25, \dodoi{10.1051/0004-6361/201936012}

\bibitem[{{Jonusas} {et~al.}(2020){Jonusas}, {Leroux}, \& {Krim}}]{Jonusas2020}
{Jonusas}, M., {Leroux}, K., \& {Krim}, L. 2020, Journal of Molecular
  Structure, 1220, 128736, \dodoi{10.1016/j.molstruc.2020.128736}

\bibitem[{{J{\o}rgensen} {et~al.}(2002){J{\o}rgensen}, {Sch{\"o}ier}, \& {van
  Dishoeck}}]{Jorgensen02}
{J{\o}rgensen}, J.~K., {Sch{\"o}ier}, F.~L., \& {van Dishoeck}, E.~F. 2002,
  \aap, 389, 908, \dodoi{10.1051/0004-6361:20020681}

\bibitem[{{J{\o}rgensen} {et~al.}(2018){J{\o}rgensen}, {M{\"u}ller}, {Calcutt},
  {Coutens}, {Drozdovskaya}, {{\"O}berg}, {Persson}, {Taquet}, {van Dishoeck},
  \& {Wampfler}}]{Jorgensen18}
{J{\o}rgensen}, J.~K., {M{\"u}ller}, H.~S.~P., {Calcutt}, H., {et~al.} 2018,
  \aap, 620, A170, \dodoi{10.1051/0004-6361/201731667}

\bibitem[{{Kristensen} {et~al.}(2012){Kristensen}, {van Dishoeck}, {Bergin},
  {Visser}, {Y{\i}ld{\i}z}, {San Jose-Garcia}, {J{\o}rgensen}, {Herczeg},
  {Johnstone}, {Wampfler}, {Benz}, {Bruderer}, {Cabrit}, {Caselli}, {Doty},
  {Harsono}, {Herpin}, {Hogerheijde}, {Karska}, {van Kempen}, {Liseau},
  {Nisini}, {Tafalla}, {van der Tak}, \& {Wyrowski}}]{Kristensen12}
{Kristensen}, L.~E., {van Dishoeck}, E.~F., {Bergin}, E.~A., {et~al.} 2012,
  \aap, 542, A8, \dodoi{10.1051/0004-6361/201118146}

\bibitem[{{Kukolich}(1967)}]{Kukolich67}
{Kukolich}, S.~G. 1967, Physical Review, 156, 83,
  \dodoi{10.1103/PhysRev.156.83}

\bibitem[{{Kukolich} \& {Wofsy}(1970)}]{Kukolich70}
{Kukolich}, S.~G., \& {Wofsy}, S.~C. 1970, \jcp, 52, 5477,
  \dodoi{10.1063/1.1672799}

\bibitem[{{Lacy} {et~al.}(1998){Lacy}, {Faraji}, {Sandford}, \&
  {Allamandola}}]{Lacy1998}
{Lacy}, J.~H., {Faraji}, H., {Sandford}, S.~A., \& {Allamandola}, L.~J. 1998,
  \apjl, 501, L105, \dodoi{10.1086/311452}

\bibitem[{{Lay} {et~al.}(1995){Lay}, {Carlstrom}, \& {Hills}}]{Lay95}
{Lay}, O.~P., {Carlstrom}, J.~E., \& {Hills}, R.~E. 1995, \apjl, 452, L73,
  \dodoi{10.1086/309713}

\bibitem[{{Le Gal} {et~al.}(2014){Le Gal}, {Hily-Blant}, {Faure}, {Pineau des
  For{\^e}ts}, {Rist}, \& {Maret}}]{LeGal2014}
{Le Gal}, R., {Hily-Blant}, P., {Faure}, A., {et~al.} 2014, \aap, 562, A83,
  \dodoi{10.1051/0004-6361/201322386}

\bibitem[{{Loinard} {et~al.}(2001){Loinard}, {Castets}, {Ceccarelli}, {Caux},
  \& {Tielens}}]{Loinard01}
{Loinard}, L., {Castets}, A., {Ceccarelli}, C., {Caux}, E., \& {Tielens},
  A.~G.~G.~M. 2001, \apjl, 552, L163, \dodoi{10.1086/320331}

\bibitem[{{Looney} {et~al.}(2000){Looney}, {Mundy}, \& {Welch}}]{Looney00}
{Looney}, L.~W., {Mundy}, L.~G., \& {Welch}, W.~J. 2000, \apj, 529, 477,
  \dodoi{10.1086/308239}

\bibitem[{{Manigand} {et~al.}(2020){Manigand}, {J{\o}rgensen}, {Calcutt},
  {M{\"u}ller}, {Ligterink}, {Coutens}, {Drozdovskaya}, {van Dishoeck}, \&
  {Wampfler}}]{Manigand20}
{Manigand}, S., {J{\o}rgensen}, J.~K., {Calcutt}, H., {et~al.} 2020, \aap, 635,
  A48, \dodoi{10.1051/0004-6361/201936299}

\bibitem[{{Maret} {et~al.}(2006){Maret}, {Bergin}, \& {Lada}}]{Maret06}
{Maret}, S., {Bergin}, E.~A., \& {Lada}, C.~J. 2006, \nat, 442, 425,
  \dodoi{10.1038/nature04919}

\bibitem[{{Mart{\'\i}n-Dom{\'e}nech} {et~al.}(2014){Mart{\'\i}n-Dom{\'e}nech},
  {Mu{\~n}oz Caro}, {Bueno}, \& {Goesmann}}]{Martin-Domenech2014}
{Mart{\'\i}n-Dom{\'e}nech}, R., {Mu{\~n}oz Caro}, G.~M., {Bueno}, J., \&
  {Goesmann}, F. 2014, \aap, 564, A8, \dodoi{10.1051/0004-6361/201322824}

\bibitem[{{McMullin} {et~al.}(2007){McMullin}, {Waters}, {Schiebel}, {Young},
  \& {Golap}}]{CASA}
{McMullin}, J.~P., {Waters}, B., {Schiebel}, D., {Young}, W., \& {Golap}, K.
  2007, in Astronomical Society of the Pacific Conference Series, Vol. 376,
  Astronomical Data Analysis Software and Systems XVI, ed. R.~A. {Shaw},
  F.~{Hill}, \& D.~J. {Bell}, 127

\bibitem[{{Millar} {et~al.}(1989){Millar}, {Bennett}, \& {Herbst}}]{Millar89}
{Millar}, T.~J., {Bennett}, A., \& {Herbst}, E. 1989, \apj, 340, 906,
  \dodoi{10.1086/167444}

\bibitem[{{Minissale} {et~al.}(2022){Minissale}, {Aikawa}, {Bergin}, {Bertin},
  {Brown}, {Cazaux}, {Charnley}, {Coutens}, {Cuppen}, {Guzman}, {Linnartz},
  {McCoustra}, {Rimola}, {Schrauwen}, {Toubin}, {Ugliengo}, {Watanabe},
  {Wakelam}, \& {Dulieu}}]{Minissale2022}
{Minissale}, M., {Aikawa}, Y., {Bergin}, E., {et~al.} 2022, ACS Earth and Space
  Chemistry, 6, 597, \dodoi{10.1021/acsearthspacechem.1c00357}

\bibitem[{{Mottram} {et~al.}(2013){Mottram}, {van Dishoeck}, {Schmalzl},
  {Kristensen}, {Visser}, {Hogerheijde}, \& {Bruderer}}]{Mottram13}
{Mottram}, J.~C., {van Dishoeck}, E.~F., {Schmalzl}, M., {et~al.} 2013, \aap,
  558, A126, \dodoi{10.1051/0004-6361/201321828}

\bibitem[{{M{\"u}ller} {et~al.}(2005){M{\"u}ller}, {Schl{\"o}der}, {Stutzki},
  \& {Winnewisser}}]{CDMS}
{M{\"u}ller}, H. S.~P., {Schl{\"o}der}, F., {Stutzki}, J., \& {Winnewisser}, G.
  2005, Journal of Molecular Structure, 742, 215,
  \dodoi{10.1016/j.molstruc.2005.01.027}

\bibitem[{{Nomura} \& {Millar}(2004)}]{nomura2004}
{Nomura}, H., \& {Millar}, T.~J. 2004, \aap, 414, 409,
  \dodoi{10.1051/0004-6361:20031646}

\bibitem[{{Notsu} {et~al.}(2021){Notsu}, {van Dishoeck}, {Walsh}, {Bosman}, \&
  {Nomura}}]{Notsu2021}
{Notsu}, S., {van Dishoeck}, E.~F., {Walsh}, C., {Bosman}, A.~D., \& {Nomura},
  H. 2021, \aap, 650, A180, \dodoi{10.1051/0004-6361/202140667}

\bibitem[{{{\"O}berg} \& {Bergin}(2021)}]{Oberg21}
{{\"O}berg}, K.~I., \& {Bergin}, E.~A. 2021, \physrep, 893, 1,
  \dodoi{10.1016/j.physrep.2020.09.004}

\bibitem[{{{\"O}berg} {et~al.}(2011){{\"O}berg}, {Boogert}, {Pontoppidan}, {van
  den Broek}, {van Dishoeck}, {Bottinelli}, {Blake}, \& {Evans}}]{Oberg11}
{{\"O}berg}, K.~I., {Boogert}, A.~C.~A., {Pontoppidan}, K.~M., {et~al.} 2011,
  \apj, 740, 109, \dodoi{10.1088/0004-637X/740/2/109}

\bibitem[{{Ortiz-Le{\'o}n} {et~al.}(2018){Ortiz-Le{\'o}n}, {Loinard}, {Dzib},
  {Galli}, {Kounkel}, {Mioduszewski}, {Rodr{\'\i}guez}, {Torres}, {Hartmann},
  {Boden}, {Evans}, {Brice{\~n}o}, \& {Tobin}}]{Ortiz-Leon18}
{Ortiz-Le{\'o}n}, G.~N., {Loinard}, L., {Dzib}, S.~A., {et~al.} 2018, \apj,
  865, 73, \dodoi{10.3847/1538-4357/aada49}

\bibitem[{{Ossenkopf} \& {Henning}(1994)}]{Ossenkopf94}
{Ossenkopf}, V., \& {Henning}, T. 1994, \aap, 291, 943

\bibitem[{{Persson} {et~al.}(2016){Persson}, {Harsono}, {Tobin}, {van
  Dishoeck}, {J{\o}rgensen}, {Murillo}, \& {Lai}}]{Persson16}
{Persson}, M.~V., {Harsono}, D., {Tobin}, J.~J., {et~al.} 2016, \aap, 590, A33,
  \dodoi{10.1051/0004-6361/201527666}

\bibitem[{{Persson} {et~al.}(2014){Persson}, {J{\o}rgensen}, {van Dishoeck}, \&
  {Harsono}}]{Persson14}
{Persson}, M.~V., {J{\o}rgensen}, J.~K., {van Dishoeck}, E.~F., \& {Harsono},
  D. 2014, \aap, 563, A74, \dodoi{10.1051/0004-6361/201322845}

\bibitem[{{Pickett} {et~al.}(1998){Pickett}, {Poynter}, {Cohen}, {Delitsky},
  {Pearson}, \& {M{\"u}ller}}]{Pickett1988}
{Pickett}, H.~M., {Poynter}, R.~L., {Cohen}, E.~A., {et~al.} 1998, \jqsrt, 60,
  883, \dodoi{10.1016/S0022-4073(98)00091-0}

\bibitem[{{Pineda} {et~al.}(2022){Pineda}, {Harju}, {Caselli}, {Sipil{\"a}},
  {Juvela}, {Vastel}, {Rosolowsky}, {Burkert}, {Friesen}, {Shirley},
  {Maureira}, {Choudhury}, {Segura-Cox}, {G{\"u}sten}, {Punanova}, {Bizzocchi},
  \& {Goodman}}]{Pineda2022}
{Pineda}, J.~E., {Harju}, J., {Caselli}, P., {et~al.} 2022, \aj, 163, 294,
  \dodoi{10.3847/1538-3881/ac6be7}

\bibitem[{{Przybilla} {et~al.}(2008){Przybilla}, {Nieva}, \&
  {Butler}}]{Przybilla08}
{Przybilla}, N., {Nieva}, M.-F., \& {Butler}, K. 2008, \apjl, 688, L103,
  \dodoi{10.1086/595618}

\bibitem[{{Rosolowsky} {et~al.}(2008){Rosolowsky}, {Pineda}, {Foster},
  {Borkin}, {Kauffmann}, {Caselli}, {Myers}, \& {Goodman}}]{Rosolowsky08}
{Rosolowsky}, E.~W., {Pineda}, J.~E., {Foster}, J.~B., {et~al.} 2008, \apjs,
  175, 509, \dodoi{10.1086/524299}

\bibitem[{{Rubin} {et~al.}(2015){Rubin}, {Altwegg}, {Balsiger}, {Bar-Nun},
  {Berthelier}, {Bieler}, {Bochsler}, {Briois}, {Calmonte}, {Combi}, {De
  Keyser}, {Dhooghe}, {Eberhardt}, {Fiethe}, {Fuselier}, {Gasc}, {Gombosi},
  {Hansen}, {H{\"a}ssig}, {J{\"a}ckel}, {Kopp}, {Korth}, {Le Roy}, {Mall},
  {Marty}, {Mousis}, {Owen}, {R{\`e}me}, {S{\'e}mon}, {Tzou}, {Waite}, \&
  {Wurz}}]{Rubin15}
{Rubin}, M., {Altwegg}, K., {Balsiger}, H., {et~al.} 2015, Science, 348, 232,
  \dodoi{10.1126/science.aaa6100}

\bibitem[{{Sahu} {et~al.}(2019){Sahu}, {Liu}, {Su}, {Li}, {Lee}, {Hirano}, \&
  {Takakuwa}}]{Su19}
{Sahu}, D., {Liu}, S.-Y., {Su}, Y.-N., {et~al.} 2019, \apj, 872, 196,
  \dodoi{10.3847/1538-4357/aaffda}

\bibitem[{{Sch{\"o}ier} {et~al.}(2005){Sch{\"o}ier}, {van der Tak}, {van
  Dishoeck}, \& {Black}}]{LAMDA}
{Sch{\"o}ier}, F.~L., {van der Tak}, F.~F.~S., {van Dishoeck}, E.~F., \&
  {Black}, J.~H. 2005, \aap, 432, 369, \dodoi{10.1051/0004-6361:20041729}

\bibitem[{{Schwarz} \& {Bergin}(2014)}]{Schwarz14}
{Schwarz}, K.~R., \& {Bergin}, E.~A. 2014, \apj, 797, 113,
  \dodoi{10.1088/0004-637X/797/2/113}

\bibitem[{{Shah} \& {Wootten}(2001)}]{Shah01}
{Shah}, R.~Y., \& {Wootten}, A. 2001, \apj, 554, 933, \dodoi{10.1086/321396}

\bibitem[{{Shirley}(2015)}]{Shirley15}
{Shirley}, Y.~L. 2015, \pasp, 127, 299, \dodoi{10.1086/680342}

\bibitem[{{Taquet} {et~al.}(2014){Taquet}, {Charnley}, \&
  {Sipil{\"a}}}]{Taquet14}
{Taquet}, V., {Charnley}, S.~B., \& {Sipil{\"a}}, O. 2014, \apj, 791, 1,
  \dodoi{10.1088/0004-637X/791/1/1}

\bibitem[{{Taquet} {et~al.}(2013){Taquet}, {L{\'o}pez-Sepulcre}, {Ceccarelli},
  {Neri}, {Kahane}, {Coutens}, \& {Vastel}}]{Taquet13}
{Taquet}, V., {L{\'o}pez-Sepulcre}, A., {Ceccarelli}, C., {et~al.} 2013, \apjl,
  768, L29, \dodoi{10.1088/2041-8205/768/2/L29}

\bibitem[{{Taquet} {et~al.}(2019){Taquet}, {Bianchi}, {Codella}, {Persson},
  {Ceccarelli}, {Cabrit}, {J{\o}rgensen}, {Kahane}, {L{\'o}pez-Sepulcre}, \&
  {Neri}}]{Taquet19}
{Taquet}, V., {Bianchi}, E., {Codella}, C., {et~al.} 2019, \aap, 632, A19,
  \dodoi{10.1051/0004-6361/201936044}

\bibitem[{{Teague} \& {Foreman-Mackey}(2018)}]{Teague18_bettermoments}
{Teague}, R., \& {Foreman-Mackey}, D. 2018, Research Notes of the American
  Astronomical Society, 2, 173, \dodoi{10.3847/2515-5172/aae265}

\bibitem[{{Tinacci} {et~al.}(2022){Tinacci}, {Germain}, {Pantaleone},
  {Ferrero}, {Ceccarelli}, \& {Ugliengo}}]{Tinacci2022}
{Tinacci}, L., {Germain}, A., {Pantaleone}, S., {et~al.} 2022, ACS Earth and
  Space Chemistry, 6, 1514, \dodoi{10.1021/acsearthspacechem.2c00040}

\bibitem[{{van der Tak} {et~al.}(2002){van der Tak}, {Schilke}, {M{\"u}ller},
  {Lis}, {Phillips}, {Gerin}, \& {Roueff}}]{vanderTak02}
{van der Tak}, F.~F.~S., {Schilke}, P., {M{\"u}ller}, H.~S.~P., {et~al.} 2002,
  \aap, 388, L53, \dodoi{10.1051/0004-6361:20020647}

\bibitem[{{van Dishoeck} {et~al.}(1993){van Dishoeck}, {Blake}, {Draine}, \&
  {Lunine}}]{VanDishoeck93}
{van Dishoeck}, E.~F., {Blake}, G.~A., {Draine}, B.~T., \& {Lunine}, J.~I.
  1993, in Protostars and Planets III, ed. E.~H. {Levy} \& J.~I. {Lunine}, 163

\bibitem[{{Wilson} {et~al.}(2009){Wilson}, {Rohlfs}, \&
  {H{\"u}ttemeister}}]{Wilson09_ToRA}
{Wilson}, T.~L., {Rohlfs}, K., \& {H{\"u}ttemeister}, S. 2009, {Tools of Radio
  Astronomy}, \dodoi{10.1007/978-3-540-85122-6}

\bibitem[{{Yu} {et~al.}(2010){Yu}, {Pearson}, {Drouin}, {Sung}, {Pirali},
  {Vervloet}, {Martin-Drumel}, {Endres}, {Shiraishi}, {Kobayashi}, \&
  {Matsushima}}]{Yu2010}
{Yu}, S., {Pearson}, J.~C., {Drouin}, B.~J., {et~al.} 2010, \jcp, 133, 174317,
  \dodoi{10.1063/1.3499911}

\bibitem[{{Zucker} {et~al.}(2020){Zucker}, {Speagle}, {Schlafly}, {Green},
  {Finkbeiner}, {Goodman}, \& {Alves}}]{Zucker20}
{Zucker}, C., {Speagle}, J.~S., {Schlafly}, E.~F., {et~al.} 2020, \aap, 633,
  A51, \dodoi{10.1051/0004-6361/201936145}

\end{thebibliography}
\bibliographystyle{aasjournal}



\end{document}